\documentclass[pdflatex,sn-mathphys-num]{sn-jnl}% Math and Physical Sciences Numbered Reference Style
\usepackage{graphicx}%
\usepackage{multirow}%
\usepackage{amsmath,amssymb,amsfonts}%
\usepackage{amsthm}%
\usepackage{mathrsfs}%
\usepackage[title]{appendix}%
\usepackage{xcolor}%
\usepackage{textcomp}%
\usepackage{manyfoot}%
\usepackage{booktabs}%
\usepackage{algorithm}%
\usepackage{algorithmicx}%
\usepackage{algpseudocode}%
\usepackage{listings}%
\usepackage{bm}
\usepackage{multirow}
\usepackage{rotating}
\usepackage{subcaption}
\usepackage{caption}

\theoremstyle{thmstyleone}%
\theoremstyle{thmstyletwo}%
\newtheorem{remark}{Remark}%

\theoremstyle{thmstylethree}%

\newcommand\dd{\mathrm{d}}

\begin{document}

\title[Robust and Sparse GLM via MMD]{Robust and Sparse Generalized Linear Models for High-Dimensional Data via Maximum Mean Discrepancy}


\author[1]{\fnm{Xiaoning} \sur{Kang}}\email{kangxiaoning@dufe.edu.cn}

\author*[2]{\fnm{Lulu} \sur{Kang}}\email{lulukang@umass.edu}

\affil[1]{\orgdiv{Institute of Supply Chain Analytics and International Business College}, \orgname{Dongbei University of Finance and Economics}, \orgaddress{\street{217 Jianshan Street}, \city{Dalian}, \postcode{116025}, \state{Liaoning}, \country{China}}}

\affil*[2]{\orgdiv{Department of Mathematics and Statistics}, \orgname{University of Massachusetts Amherst}, \orgaddress{\street{710 N Pleasant St.}, \city{Amherst}, \postcode{01003}, \state{MA}, \country{U.S.A.}}}


\abstract{High-dimensional datasets are frequently subject to contamination by outliers and heavy-tailed noise, which can severely bias standard regularized estimators like the Lasso. While Maximum Mean Discrepancy (MMD) has recently been introduced as a ``universal'' framework for robust regression, its application to high-dimensional Generalized Linear Models (GLMs) remains largely unexplored, particularly regarding variable selection. 
In this paper, we propose a penalized MMD framework for robust estimation and feature selection in GLMs. We introduce an $\ell_1$-penalized MMD objective and develop two versions of the estimator: a full $O(n^2)$ version and a computationally efficient $O(n)$ approximation. 
To solve the resulting non-convex optimization problem, we employ an algorithm based on the Alternating Direction Method of Multipliers (ADMM) combined with AdaGrad. 
Through extensive simulation studies involving Gaussian linear regression and binary logistic regression, we demonstrate that our proposed methods are highly competitive with classical penalized GLMs and existing robust benchmarks. 
Our approach shows particular resilience in maintaining a balance between estimation accuracy and variable selection across diverse contamination scenarios, especially in handling high-leverage points and heavy-tailed error distributions where traditional methods may fluctuate in performance.}


\maketitle

\section{Introduction}\label{sec:intro}
The proliferation of high-dimensional datasets in fields ranging from genomics to finance has necessitated the development of statistical models that are both sparse and resilient. 
In the context of Generalized Linear Models (GLMs), standard estimation techniques like Maximum Likelihood Estimation (MLE) are sensitive to even minor data contamination or the presence of heavy-tailed noise. 
While the introduction of shrinkage penalties—most notably the Lasso ($\ell_1$) and Ridge ($\ell_2$) regularizers—has allowed for effective variable selection and stable estimation in high-dimensional settings where $p \gg n$ with $p$ denotating the dimension of input variables and $n$ sample size, these classical penalized estimators generally inherit the lack of robustness of their underlying loss functions.

Robustness in regression has traditionally been addressed through M-estimators (e.g. Huber loss by \cite{huber1964robust}) and have been extended to generalized linear models combined with variable selection or regularization methods including \cite{khan2007robust}, \cite{alfons2016robust}, \cite{loh2017statistical}, \cite{chang2018robust}, etc. 
Another group of robust estimators were also introduced based on the use of density power divergence (DPD) \citep{basu1998robust}, such as the DPD-based linear regression \citep{ghosh2013robust} and $L_2E$ criterion for classification \citep{chi2014robust}. 
However, many of these methods are either tailored to specific distribution families or face significant optimization challenges in non-convex landscapes. 
Recently, Maximum Mean Discrepancy (MMD) has emerged as a powerful ``universal'' framework for robust inference. 
By treating estimation as a minimum distance problem in a Reproducing Kernel Hilbert Space (RKHS), MMD-based estimators can achieve strong robustness properties—specifically bounded influence functions—without requiring the model to be well-specified \citep{alquier2024universal}.

While \cite{alquier2024universal} established the theoretical consistency and robustness of MMD estimators for arbitrary regression models, their framework was primarily designed for low-to-moderate dimensional settings. 
Specifically, their approach does not incorporate the shrinkage mechanisms necessary to handle the sparsity constraints common in modern big data applications. 
In regimes with a large number of predictors (including high-dimensional cases where $p \gg n$), an unpenalized MMD estimator may suffer from over-fitting or failure of identifiability, limiting its practical utility for feature selection.

In this paper, we fill this gap by proposing a penalized MMD framework for GLMs. 
The original MMD-based estimators have been discussed in the context of ``universal'' robustness \cite{alquier2024universal}. 
Our work provides a novel bridge to high-dimensional statistics. 
The originality of this research lies not merely in applying MMD to regression, but in the formulation of penalized MMD-GLM objectives and the demonstration that MMD’s joint-distribution perspective offers a unique solution to the ``leverage point'' problem in high dimensions -- a scenario where traditional robust M-estimators often fail due to their reliance on coordinate-wise bounded influence functions.
Second, we develop an efficient optimization algorithm based on the Alternating Direction Method of Multipliers (ADMM) and AdaGrad to solve the resulting non-convex regularized problems. 
Third, we provide numerical evidence across binary classification and linear regression tasks, demonstrating that our proposed estimator maintains high predictive accuracy even under significant adversarial contamination, where classical penalized GLMs and unpenalized robust methods fail.

The remainder of this paper is organized as follows: Section 2 provides a brief preliminary on MMD and its application to GLMs. Section 3 introduces our proposed penalized research framework and the specific formulations for classification and regression. Section 4 details our computational strategy. Section 5 and 6 present simulation results and examples based on real data comparing our method to existing benchmarks, and Section 7 concludes with a discussion of future directions.

\section{Preliminaries and MMD-based GLM Estimators}\label{sec:background}

In this section, we review the background on MMD and two different versions of the robust GLM estimators proposed in \cite{alquier2024universal}. 
Let $\mathcal{Z}$ be a topological space equipped with a Borel $\sigma$-algebra $\mathcal{S}_Z$. 
We consider a symmetric, positive definite kernel $K: \mathcal{Z} \times \mathcal{Z} \to \mathbb{R}$ and its associated Reproducing Kernel Hilbert Space (RKHS), denoted by $\mathcal{H}_K$. 
Through this work, we assume that the kernel $K$ is $\mathfrak{S}_{\mathcal{Z}}$-measurable and bounded, such that $\sup_{z \in \mathcal{Z}} K(z,z) < \infty$. 

The Maximum Mean Discrepancy (MMD) \citep{gretton2012kernel} is a distance between probability distributions defined by the embedding of these distributions into the RKHS. 
For any two probability distributions $P_1$ and $P_2$ in the set of all probability distributions $\mathcal{P}(\mathcal{Z})$, the squared MMD distance is defined as:
\begin{equation}\label{eq:mmd}
D(K,\mathcal{Z}, P_1,P_2)^2=\mathbb{E}_{\bm z, \bm z'\sim P_1}[K(\bm z,\bm z')]-2\mathbb{E}_{\bm z\sim P_1, \bm t\sim P_2}[K(\bm z,\bm t)]+\mathbb{E}_{\bm t, \bm t'\sim P_2}[K(\bm t, \bm t')]
\end{equation}
When the kernel $K$ is a \emph{characteristic kernel}, such as Gaussian or Mat\'{e}rn kernel, $D(K, \mathcal{Z}, P_1, P_2) = 0$ if and only if $P_1 = P_2$, making it a proper metric on the space of probability distributions.
We omit more detailed background on kernel (conditional) mean embedding. 
Readers can found them in \cite{gretton2012kernel} and \cite{fukumizu2004dimensionality}, and how they are applied for GLM in \cite{alquier2024universal}.

In the context of supervised learning, we define the product space $\mathcal{Z} = \mathcal{X} \times \mathcal{Y}$, where $\mathcal{X}\subset \mathbb{R}^p$ and $\mathcal{Y}\subset \mathbb{R}$ are two topological spaces, equipped respectively with the Borel $\sigma$-algebra $\mathfrak{S}_{\mathcal{X}}$ and $\mathfrak{S}_{\mathcal{Y}}$.
The space $\mathcal{X}$ is for the $p$-dimensional input variables $\bm X$ and $\mathcal{Y}$ for the univariate response variable $Y$.
The observed data $D_n=\{\bm x_i, y_i\}$ are $n$ i.i.d. samples for the random variable $\bm Z=(\bm X,Y) \in \mathcal{Z}=\mathcal{X}\times \mathcal{Y}$. 
Therefore, $\mathcal{Z}$ is equipped with the $\sigma$-algebra $\mathfrak{S}_{\mathcal{Z}}=\mathfrak{S}_{\mathcal{X}}\otimes \mathfrak{S}_{\mathcal{Y}}$.
We denote $P(\mathcal{Z})$ as the set of all probability distributions on $(\mathcal{Z}, \mathfrak{S}_{\mathcal{Z}})$.
To focus on GLM, we make a parametric assumption on the distribution of $\mathcal{Y}$.
Let $\{P_{\lambda}, \lambda \in \Lambda \}$ be a set of probability distributions on $\mathcal{Y}$, $\Theta$ be the space of parameters $\bm \theta$, and $g$ is the link function $g: \Theta \times \mathcal{X} \rightarrow \Lambda$ such that the mapping $\bm x \mapsto P_{\lambda=g(\bm x,\bm \theta)}(A)$ is $\mathfrak{S}_{\mathcal{Y}}$-measurable for all $A \in \mathfrak{S}_{\mathcal{Y}}$ and all $\bm \theta \in \Theta$.
So the distribution of $Y_i$ given $\bm X=\bm x_i$ is denoted by $P_{g(\bm \theta,\bm x_i)}$ for $\bm \theta \in \Theta$.
Assume $P_{\lambda}$ has a density $p_{\lambda}$ and both $p_{\lambda}$ and $g(\bm \theta, \bm x)$ is differentiable with respect to both $\bm x$ and $\bm \theta$.
We assume $K$ can be factorized into two $K_{x}$ and $K_y$, where $K_{x}$ and $K_y$ are symmetric and positive definite kernels on $\mathcal{X}$ and $\mathcal{Y}$, respectively.
For any $\bm z_1=(\bm x_1, y_1), \bm z_2=(\bm x_2, y_2) \in \mathcal{Z}$,
\[
K(\bm z_1, \bm z_2)=K_x(\bm x_1, \bm x_2)K_y(y_1,y_2).
\]
What is more, we assume both $K_x$ and $K_y$ are finitely bounded.
If $\bm X$ are all continuous input variables, we then assume $K_x$ is continuous as well.

Given the data $D_n$, denote by $\hat{P}^{n}$ the empirical distribution with density function $\hat{p}^{n}=\frac{1}{n}\sum_{i=1}^n \delta_{\bm z_i}(\bm z)$ where $\bm z_i=(\bm x_i, y_i)$ and $\delta_{\bm z_i}(\bm z)=1$ if $\bm z=\bm z_i$ and 0 otherwise.
Similarly, we can also define $\hat{P}^n_x$ as the empirical distribution for $\bm X$ on the set of $\{\bm x_1,\ldots, \bm x_n\}$.
The parametric distribution for the data based on the GLM assumption is defined as 
\[
P^{n}_{\bm \theta}(A\times B)=\frac{1}{n} \sum_{i=1}^n \delta_{\bm x_i}(A)P_{g(\bm \theta, x_i)}(B), \quad A \in \mathfrak{S}_{\mathcal{X}}, B\in \mathfrak{S}_{\mathcal{Y}}.
\]
Essentially, $P^{n}_{\bm \theta}$ assumes a discrete uniform distribution on the set $\{\bm x_1,\ldots, \bm x_n\}$ and $Y|\bm x_i=P_{g(\bm \theta, \bm x_i)}$ is the GLM.

Based on the general definition of MMD, we derive the squared MMD for the GLM given data.
\begin{align*}
&D(K,\mathcal{Z}, \hat{P}^n,P^n_{\bm \theta})^2\\
=&\mathbb{E}_{\bm Z, \bm Z'\sim P^n_{\bm \theta}}[K(\bm Z,\bm Z')]-2\mathbb{E}_{\bm Z\sim P^n_{\bm \theta}, \bm T\sim \hat{P}^n}[K(\bm Z,\bm T)]+\mathbb{E}_{\bm T, \bm T'\sim \hat{P}^n}[K(\bm T, \bm T')]\\
=&\textcircled{1}+\textcircled{2}+\textcircled{3}.
\end{align*}
Since $\hat{P}^n$ is the empirical distribution, $\textcircled{3}=\frac{1}{n^2}\sum_{i,j=1}^n K(\bm z_i,\bm z_j)$ does not depend on $\bm \theta$.
The terms $\textcircled{1}$ and $\textcircled{2}$ can be derived as follows.
Readers can find detailed derivation in the Supplementary Material. 
\begin{align*}
\textcircled{1}&=\frac{1}{n^2}\sum_{i,j=1}^n K_x(\bm x_i, \bm x_j)\mathbb{E}_{Y\sim P_{g(\bm \theta,\bm x_i)}, Y'\sim P_{g(\bm \theta, \bm x_j)}}[K_y(Y,Y')],\\
\textcircled{2}&=-\frac{2}{n^2}\sum_{i,j=1}^n K_x(\bm x_i,\bm x_j) \int K_y(y, y_i)p_{g(\bm \theta, \bm x_j)}(y)\dd y.
\end{align*}
The part of $D(K,\mathcal{Z}, \hat{P}^n,P^n_{\bm \theta})^2$ involving $\bm \theta$ is $\textcircled{1}+\textcircled{2}$ and can be simplified as 
\begin{align*}
\textcircled{1}+\textcircled{2}&= \frac{1}{n^2}\sum_{i,j}^n K_x(\bm x_i,\bm x_j)l(\bm \theta, \bm x_i, \bm x_j, y_i),
\end{align*}
where
\begin{align*}
l(\bm \theta, \bm x_i, \bm x_j, y_i)&:=\int K_y(y, y') p_{g(\bm \theta, \bm x_i)}(y) p_{g(\bm \theta, \bm x_j)}(y') \dd y \dd y'-2\int K_y(y, y_i) p_{g(\bm \theta,\bm x_j)}(y)\dd y\\
&=\mathbb{E}_{Y\sim P_{g(\bm \theta,\bm x_i)}, Y'\sim P_{g(\bm \theta, \bm x_j)}}[K_y(Y,Y')]-2\mathbb{E}_{Y\sim P_{g(\bm \theta,\bm x_j)}}[K_y(Y,y_i)]
\end{align*}
In \cite{alquier2024universal}, the authors defined $\hat{\bm \theta}_n$ in the following as the robust GLM estimator based on MMD.
\[
\hat{\bm \theta}_n \in \text{arg}\min_{\bm \theta} \sum_{i,j=1}^n K_x(\bm x_i,\bm x_j)l(\bm \theta, \bm x_i, \bm x_j, y_i).
\]

To compute the objective function, it requires $O(n^2)$ order of computation for $l(\bm \theta, \bm x_i, \bm x_j, y_i)$.
In \cite{alquier2024universal}, the authors further decomposed $\textcircled{1}+\textcircled{2}$ into the two groups: $i=j$ group and $i\neq j$ group, i.e., 
\begin{align*}
n^2\times [\textcircled{1}+\textcircled{2}]&= \sum_{i=1}^n K_x(\bm x_i,\bm x_i)l(\bm \theta, \bm x_i, \bm x_i, y_i)+\sum_{i\neq j}^n K_x(\bm x_i, \bm x_j)l(\bm \theta, \bm x_i, \bm x_j, y_i)\\
&=K_x(\bm x, \bm x)\sum_{i=1}^n \tilde{l}(\bm \theta, \bm x_i, y_i)+\sum_{i\neq j}^n K_x(\bm x_i, \bm x_j)l(\bm \theta, \bm x_i, \bm x_j, y_i),
\end{align*}
where
\begin{align*}
\tilde{l}(\bm \theta, \bm x_i, y_i)&=\int K_y(y, y') p_{g(\bm \theta, \bm x_i)}(y) p_{g(\bm \theta, \bm x_i)}(y') \dd y \dd y'-2\int K_y(y, y_i) p_{g(\bm \theta,\bm x_i)}(y)\dd y\\
&=\mathbb{E}_{Y,Y'\sim P_{g(\bm \theta, \bm x_i)}}[K_y(Y,Y')]-2\mathbb{E}_{Y\sim P_{g(\bm \theta, \bm x_i)}}[K_y(y_i, Y)].
\end{align*}
If we assume the kernel function $K_x$ decreases quickly as $||\bm x_i-\bm x_j||$ increases, for example, $K_x(\bm x_i, \bm x_j)=\exp(-\frac{1}{h}||\bm x_i-\bm x_j||^2)$ with very small $h$, then the cross term with $i\neq j$ can be small or even negligible, especially for large dimension $p$.
Therefore, $\tilde{\bm \theta}_n$ can be an approximate to $\hat{\bm \theta}_n$ and it only requires $O(n)$ order of computation.
\[
\tilde{\bm \theta}_n \in \text{arg}\min_{\Theta} \sum_{i=1}^n l(\bm \theta, \bm x_i,y_i).
\]
In fact, under some assumptions, including the ones on the kernel function mentioned above, $\hat{\bm \theta}_n\rightarrow \tilde{\bm \theta}_n$ as the bandwidth goes to 0. 
Both $\hat{\bm \theta}_n$ and $\tilde{\bm \theta}_n$ are robust estimators for $\bm \theta$.
Theoretical properties were also developed in \cite{alquier2024universal} for the two estimators. 
Under a series of assumptions, the estimator $\hat{\bm \theta}_n$ is consistent. 
Its robustness to adversarial contaminations was confirmed by the non-asymptotic error bounds under both fixed design and random design. 
The estimator $\tilde{\bm \theta}_n$ is robust to Huber-type contaminations. 

\section{Penalized MMD Estimators for Gaussian and Binary Responses}\label{sec:prop}

In high-dimensional settings where the number of predictors $p$ is large, the estimators $\hat{\bm \theta}_n$ and $\tilde{\bm \theta}_n$ introduced by \cite{alquier2024universal} may suffer from over-fitting or lack of identifiability. 
To address this, we propose adding a shrinkage penalty to the MMD objective function. 
This allows for simultaneous robust parameter estimation and variable selection.
We define the penalized $O(n^2)$ estimator as:
\begin{equation}\label{eq:v1}
\hat{\bm \theta}_n \in \arg \min_{\bm \theta \in \Theta} \sum_{i,j=1}^n K_x(\bm x_i, \bm x_j)l(\bm \theta, \bm x_i, \bm x_j, y_i) + \lambda \|\theta\|_1
\end{equation}
and the penalized $O(n)$ approximate estimator as:
\begin{equation}\label{eq:v2}
\tilde{\bm \theta}_n \in \arg \min_{\bm \theta \in \Theta} \sum_{i=1}^n \tilde{l}(\bm \theta, \bm x_i, y_i) + \lambda \|\theta\|_1
\end{equation}
where $\lambda \geq 0$ is a tuning parameter controlling the sparsity of the estimate.
In this section, we apply the general penalized estimator to the Gaussian linear regression model and the binary logistic regression model. 

\emph{Remark 1} (Statistical Interpretation and Robustness). Unlike penalized M-estimators (e.g., Huber-Lasso) which focus on the conditional distribution $P(Y|X)$ by bounding the influence of residuals, the MMD objective is a \emph{joint-distribution matching criterion} that minimizes a distance between $P(X, Y)$ and the model $P_{\bm \theta}(X, Y)$ in an RKHS. 
Conceptually, this allows the estimator to handle outliers in both the response and the covariate space (leverage points) simultaneously. 
In high-dimensional settings, the MMD loss effectively 'shields' the $\ell_1$ penalty; by providing a bounded gradient even in the presence of heavy-tailed contamination, it prevents the KKT conditions of the Lasso from being dominated by outliers, thereby maintaining the integrity of the variable selection process.

First, we consider the linear regression model where $Y(\bm x) \sim \mathcal{N}(\bm x^\top \bm \theta, \sigma^2)$. 
The density of $Y$ is $p_{g(\bm \theta, \bm x)}(y)=\frac{1}{\sqrt{2\pi}\sigma}\exp\left(-\frac{(y-\bm x^\top \bm \theta)^2}{2\sigma^2}\right)$. 
We choose the Gaussian kernel for the response variable $K_y(y_1, y_2) = \exp\left( -\frac{(y_1-y_2)^2}{2h_y^2} \right)$ because it shares the same function definition as the normal density function. 
As a result, the formulas of $l(\bm \theta, \bm x_i, \bm x_i, y_i)$ and $\tilde{l}(\bm \theta, \bm x_i, y_i)$ can be much simplified as follows. 
\begin{align}\label{eq:linearv1}
l(\bm \theta, \bm x_i, \bm x_j,  y_i)&=\frac{h_y}{\sqrt{2\sigma^2+h_y^2}}\exp\left(-\frac{1}{2}\frac{\bm \theta^\top (\bm x_j-\bm x_i)(\bm x_j-\bm x_i)^\top\bm \theta }{2\sigma^2+h_y^2}\right)\\\nonumber
&-\frac{2h_y}{\sqrt{\sigma^2+h_y^2}}\exp\left(-\frac{1}{2}\frac{(y_i-\bm x_j^\top \bm \theta)^2}{\sigma^2+h_y^2}\right),\\
\label{eq:linearv2}
\tilde{l}(\bm \theta, \bm x_i, y_i)&=\frac{h_y}{\sqrt{2\sigma^2+h_y^2}}-\frac{2h_y}{\sqrt{\sigma^2+h_y^2}}\exp\left(-\frac{1}{2}\frac{(y_i-\bm x_i^\top \bm \theta)^2}{\sigma^2+h_y^2}\right). 
\end{align}
Plugging \eqref{eq:linearv1} and \eqref{eq:linearv2} in \eqref{eq:v1} and \eqref{eq:v2}, respectively, we can obtain two versions of the MMD estimators by solving the two minimization problems. 

Second, we consider the binary classification problem. 
Assume $Y(\bm x)$ follows $\text{Bernoulli}(\pi(\bm x))$ distribution.
We choose the geometric kernel for $Y$, i.e., $K_y(y_1,y_2)=0.5h_y(1-h_y)^{|y_1-y_2|}$ for $y_1, y_2\in \{0,1\}$, which was proposed by \citep{rajagopalan1995kernel} to construct the kernel estimator for discrete distribution. 
Then the probability function for $Y$ is
\[
p_{g(\bm \theta, \bm x)}(y)=\left(\frac{e^{\bm x^\top \bm \theta}}{1+e^{\bm x^\top \bm \theta}}\right)^{y}\left(\frac{1}{1+e^{\bm x^\top \bm \theta}}\right)^{1-y}.
\]
Here
\[
\pi(\bm \theta, \bm x)=\frac{e^{\bm x^\top \bm \theta}}{1+e^{\bm x^\top \bm \theta}} \text{ and } \log \frac{\pi}{1-\pi}=\bm x^\top \bm \theta.
\]
The two function $l(\bm \theta, \bm x_i,\bm x_j,y_i)$ and $\tilde{l}(\bm \theta, \bm x_i,y_i)$ are derived as follows. 
\begin{align}\nonumber
l(\bm \theta, \bm x_i,\bm x_j,y_i)&=0.5h_y\left[1-h_y(\pi_i+\pi_j)+2h_y\pi_i\pi_j\right]\\\label{eq:logisticv1}
&-h_y\left[(1-h_y)^{1-y_i}\pi_j+(1-h_y)^{y_i}(1-\pi_j)\right],\\\label{eq:logisticv2}
\tilde{l}(\bm \theta, \bm x_i,y_i)&=h_y^2\pi_i^{2(1-y_i)}(1-\pi_i)^{2y_i}-0.5h_y. 
\end{align}
Replacing $l(\bm \theta, \bm x_i,\bm x_j,y_i)$ and $\tilde{l}(\bm \theta, \bm x_i,y_i)$ in \eqref{eq:v1} and \eqref{eq:v2} by \eqref{eq:logisticv1} and \eqref{eq:logisticv2}, respectively, we obtain the two estimators for the logistic model. 

%\cite{alquier2024universal} discussed how to compute the derivative of $l(\bm \theta, \bm x_i, \bm x_j, y_i)$ and $l(\bm \theta, \bm x_i, y_i)$.
%In general case, it is not likely to be able to compute the two exactly.
%\cite{alquier2024universal} also discussed how to estimate $l(\bm \theta, \bm x_i, \bm x_j, y_i)$ and $l(\bm \theta, \bm x_i, y_i)$.

\section{Optimization Method} 

To solve the penalized minimization problems defined in \eqref{eq:v1} and \eqref{eq:v2}, we employ the Alternating Direction Method of Multipliers (ADMM) approach \citep{boyd2011distributed}. 
This framework allows us to decouple the non-smooth shrinkage penalty from the MMD-based loss function. 
We first focus on the $O(n)$ approximate estimator with an $\ell_1$ penalty. 
The optimization problem is reformulated by introducing an auxiliary variable $\bm{\eta}$:
\begin{equation*}
\min_{\bm{\theta} \in \Theta} \frac{1}{n} \sum_{i=1}^n \tilde{l}(\bm{\theta}, \bm{x}_i, y_i) + \lambda \|\bm{\eta}\|_1 \quad \text{subject to } \bm{\theta} = \bm{\eta}.
\end{equation*}
The corresponding augmented Lagrangian is given by:
\begin{equation}\label{eq:Lv2}
\mathcal{L}(\bm{\theta}, \bm{\eta}, \bm{\gamma}) = \frac{1}{n} \sum_{i=1}^n \tilde{l}(\bm{\theta}, \bm{x}_i, y_i) + \lambda \|\bm{\eta}\|_1 + \bm{\gamma}^\top (\bm{\theta} - \bm{\eta}) + \frac{\rho}{2} \|\bm{\theta} - \bm{\eta}\|_2^2,
\end{equation}
where $\bm{\gamma}$ is the vector of Lagrange multipliers and $\rho > 0$ is a penalty parameter.
The procedure can be generalized to the $O(n^2)$ case by simply replacing $\tilde{l}(\bm{\theta}, \bm{x}_i, y_i)$ by $l(\bm \theta, \bm x_i, \bm x_j,y_i)$ and the augmented Lagrangian is
\begin{equation}\label{eq:Lv1}
\mathcal{L}(\bm{\theta}, \bm{\eta}, \bm{\gamma}) = \frac{1}{n} \sum_{i=1}^n l(\bm \theta, \bm x_i, \bm x_j,y_i) + \lambda \|\bm{\eta}\|_1 + {\bm \gamma}^\top (\bm{\theta} - \bm{\eta}) + \frac{\rho}{2} \|\bm{\theta} - \bm{\eta}\|_2^2. 
\end{equation}
Although ADMM and AdaGrad are established optimization frameworks, their application to the non-convex, kernel-based MMD objective is a significant methodological contribution. This combination allows us to decouple the global distance properties of MMD from the local sparsity constraints of the $\ell_1$ penalty, a task that standard gradient descent or coordinate descent methods struggle to perform reliably in $p \gg n$ settings.

\subsection{Gaussian Linear Regression}

In the context of linear regression, the loss $\tilde{l}(\bm{\theta}, \bm{x}_i, y_i)$ is defined as in \eqref{eq:linearv2}. 
The ADMM procedure involves the following sequential updates for iteration $k+1$:
\begin{enumerate}
\item \textbf{$\bm{\eta}$-step:} The update for the auxiliary variable $\bm{\eta}$ is found by solving $\nabla_{\bm{\eta}} \mathcal{L} = 0$. From the Lagrangian, we have:
\begin{equation*}
\frac{\partial \mathcal{L}}{\partial \bm{\eta}} = \lambda \text{sign}(\bm{\eta}) - \bm{\gamma}^{(k)} + \rho(\bm{\eta} - \bm{\theta}^{(k)}) = 0.
\end{equation*}
Rearranging this expression yields $\bm{\eta} = \bm{\theta}^{(k)} + \bm{\gamma}^{(k)}/\rho - (\lambda/\rho) \text{sign}(\bm{\eta})$, which corresponds to the element-wise soft-thresholding operator $S$:
\begin{equation*}
\bm{\eta}^{(k+1)} = S\left( \bm{\theta}^{(k)} + \frac{\bm{\gamma}^{(k)}}{\rho}, \frac{\lambda}{\rho} \right),
\end{equation*}
where $S(a, b) = \text{sign}(a) \max(|a| - b, 0)$. This step explicitly handles the sparsity of the parameter estimate.
    
\item \textbf{$\bm{\theta}$-step:} The update for $\bm{\theta}$ requires minimizing $\mathcal{L}$ with respect to $\bm{\theta}$ given $\bm{\eta}^{(k+1)}$ and $\bm{\gamma}^{(k)}$. As the MMD loss function for linear regression is only locally convex (specifically when $(y_i - \bm{x}_i^\top \bm{\theta})^2 < \sigma^2 + h_y^2$), standard gradient descent may be unstable. Following \cite{alquier2024universal}, we utilize the AdaGrad algorithm \citep{duchi2011adaptive}, which adaptively scales the learning rate for each parameter. The gradient used in this step is:
\begin{equation*}
\nabla_{\bm{\theta}} \mathcal{L} = \frac{1}{n} \sum_{i=1}^n \frac{\partial \tilde{l}_i}{\partial \bm{\theta}} + \bm{\gamma}^{(k)} + \rho(\bm{\theta} - \bm{\eta}^{(k+1)}),
\end{equation*}
where the partial derivative of the loss is:
\begin{equation*}
\frac{\partial \tilde{l}_i}{\partial \bm{\theta}} = -\frac{2h_y}{(\sigma^2 + h_y^2)^{3/2}} (y_i - \bm{x}_i^\top \bm{\theta}) \exp \left( -\frac{1}{2} \frac{(y_i - \bm{x}_i^\top \bm{\theta})^2}{\sigma^2 + h_y^2} \right) \bm{x}_i.
\end{equation*} 
     To ensure a robust start, we initialize $\bm{\theta}$ using the standard Lasso estimate and set the initial variance $\sigma^2$ based on the Lasso residuals.
\begin{remark}(Convexity of the Gaussian MMD Loss)
    The Hessian matrix of $\tilde{l}_i$ is 
    \begin{equation*}
\nabla_{\bm{\theta}}^2 \tilde{l}_i = \frac{2h_y}{(\sigma^2 + h_y^2)^{3/2}} \exp \left( -\frac{(y_i - \bm{x}_i^\top \bm{\theta})^2}{2(\sigma^2 + h_y^2)} \right) \left[ 1 - \frac{(y_i - \bm{x}_i^\top \bm{\theta})^2}{\sigma^2 + h_y^2} \right] \bm{x}_i \bm{x}_i^\top. 
\end{equation*}
Since the outer product $\bm{x}_i \bm{x}_i^\top$ is a positive semi-definite (PSD) matrix, the Hessian $\nabla_{\bm{\theta}}^2 \tilde{l}_i$ is PSD if and only if the bracketed scalar term is non-negative:
\begin{equation*}
1 - \frac{(y_i - \bm{x}_i^\top \bm{\theta})^2}{\sigma^2 + h_y^2} \geq 0 \implies (y_i - \bm{x}_i^\top \bm{\theta})^2 \leq \sigma^2 + h_y^2.
\end{equation*}
This condition implies that the loss function $\tilde{l}_i$ is locally convex in the region where the squared residual $(y_i - \bm{x}_i^\top \bm{\theta})^2$ does not exceed the sum of the model variance $\sigma^2$ and the kernel bandwidth parameter $h_y^2$. Because $\mathbb{E}[(y_i - \bm{x}_i^\top \bm{\theta})^2] = \sigma^2$, this condition is typically satisfied in the vicinity of the true parameter value, facilitating stable convergence of gradient-based optimization methods.
\end{remark}    
\item \textbf{$\sigma^2$-step:} The variance is updated in each iteration to maintain the consistency of the Gaussian kernel scale:
\begin{equation*}
(\sigma^2)^{(k+1)} = \frac{1}{n} \sum_{i=1}^n (y_i - \bm{x}_i^\top \bm{\theta}^{(k+1)})^2.
\end{equation*}

\item \textbf{$\bm{\gamma}$-step:} The dual variable is updated to enforce the constraint $\bm{\theta} = \bm{\eta}$. 
Following the ADMM approach \citep{boyd2011distributed}, gradient ascending of the dual function leads to:
\begin{equation*}
\bm{\gamma}^{(k+1)} = \bm{\gamma}^{(k)} + \rho(\bm{\theta}^{(k+1)} - \bm{\eta}^{(k+1)}).
\end{equation*}
\end{enumerate}
The algorithm terminates once the primal and dual residuals satisfy $\|\bm{\theta}^{(k+1)} - \bm{\eta}^{(k+1)}\|_2 \leq \epsilon_{\text{pri}}$ and $\|\rho(\bm{\eta}^{(k+1)} - \bm{\eta}^{(k)})\|_2 \leq \epsilon_{\text{dual}}$, respectively, where $\epsilon_{\text{pri}}$ and $\epsilon_{\text{dual}}$ are two pre-specified tolerance parameters. 

For the full $O(n^2)$ estimator $\hat{\bm{\theta}}_n$, the gradient involves the weighted sum of pairwise interactions as defined in \eqref{eq:linearv1}:
\begin{equation*}
\nabla_{\bm{\theta}} \sum_{i,j=1}^n K_x(\bm{x}_i, \bm{x}_j) l(\bm{\theta}, \bm{x}_i, \bm{x}_j, y_i) = \sum_{i,j=1}^n K_x(\bm{x}_i, \bm{x}_j) \left( \nabla_{\bm{\theta}} \text{Term 1} - \nabla_{\bm{\theta}} \text{Term 2} \right),
\end{equation*}
where, following the derivation in our notes:
\begin{align*}
\nabla_{\bm{\theta}} \text{Term 1} &= -\frac{h_y}{(2\sigma^2 + h_y^2)^{3/2}} \left[ \bm{\theta}^\top (\bm{x}_j - \bm{x}_i) \right] \exp \left( -\frac{1}{2} \frac{\bm{\theta}^\top (\bm{x}_j - \bm{x}_i)(\bm{x}_j - \bm{x}_i)^\top \bm{\theta}}{2\sigma^2 + h_y^2} \right) (\bm{x}_j - \bm{x}_i), \\
\nabla_{\bm{\theta}} \text{Term 2} &= \frac{2h_y}{(\sigma^2 + h_y^2)^{3/2}} (y_i - \bm{x}_j^\top \bm{\theta}) \exp \left( -\frac{1}{2} \frac{(y_i - \bm{x}_j^\top \bm{\theta})^2}{\sigma^2 + h_y^2} \right) \bm{x}_j.
\end{align*}
As the $O(n^2)$ gradient requires $O(n^2)$ operations per iteration, it is primarily recommended for datasets of moderate size.

\subsection{Binary Logistic Regression}

For the logistic regression model, the updates for the sparse auxiliary variable $\bm{\eta}$ and the dual variable $\bm{\gamma}$ follow the same soft-thresholding and dual-update steps as described in the linear regression section. 
The primary difference lies in the update for $\bm{\theta}$, where we utilize AdaGrad to handle the potentially non-convex MMD loss.

The gradient for the augmented Lagrangian is given by:
\begin{equation*}
\nabla_{\bm{\theta}} \mathcal{L} = \nabla_{\bm{\theta}} \left( \frac{1}{n} \sum_{i=1}^n \tilde{l}(\bm{\theta}, \bm{x}_i, y_i) \right) + \bm{\gamma}^{(k)} + \rho(\bm{\theta} - \bm{\eta}^{(k+1)}),
\end{equation*}
where the gradient of the $O(n)$ approximate loss is:
\begin{equation*}
\nabla_{\bm{\theta}} \sum_{i=1}^n \tilde{l}(\bm{\theta}, \bm{x}_i, y_i) = \frac{2}{n} \left( \sum_{y_i=0} \frac{\exp(2\bm{x}_i^\top \bm{\theta})}{(1 + \exp(\bm{x}_i^\top \bm{\theta}))^3} \bm{x}_i - \sum_{y_i=1} \frac{\exp(\bm{x}_i^\top \bm{\theta})}{(1 + \exp(\bm{x}_i^\top \bm{\theta}))^3} \bm{x}_i \right).
\end{equation*}

\begin{remark}[Convexity of the Logistic MMD Loss]
The $O(n)$ approximate loss $\tilde{l}(\bm{\theta}, \bm{x}_i, y_i)$ is defined as $\pi_i^{2(1-y_i)}(1-\pi_i)^{2y_i}$ (up to constant shifts and scales). Through the analysis of the Hessian matrix $\nabla_{\bm{\theta}}^2 \tilde{l}_i$, it can be shown that the function is not globally convex due to the properties of the sigmoid link $\pi_i = \sigma(\bm{x}_i^\top \bm{\theta})$. However, the objective function is locally convex if the following constraints are satisfied for all $i = 1, \dots, n$:
\begin{equation*}
1 - \left(\frac{2}{3}\right)^{y_i} \leq \pi(\bm{x}_i, \bm{\theta}) \leq \left(\frac{2}{3}\right)^{1-y_i}.
\end{equation*}
This interval corresponds to $\pi_i \in [0, 2/3]$ when $y_i=0$ and $\pi_i \in [1/3, 1]$ when $y_i=1$. Under these constraints, the minimization of the MMD loss becomes a convex programming problem. 
In practice, the robustness of the MMD objective combined with an initialization at the standard Lasso or logistic Lasso estimate typically places $\bm{\theta}$ within a region of the parameter space that satisfies these conditions, ensuring stable convergence.
However, if the local convexity is not met and the covergence cannot be achieved within reasonable number of iterations, we should restart the algorithm with multiple trials of intial values.
\end{remark}

For the full $O(n^2)$ estimator $\hat{\bm{\theta}}_n$, the gradient involves pairwise terms weighted by the input kernel $K_x(\bm{x}_i, \bm{x}_j)$. Using the chain rule $\nabla_{\bm{\theta}} \pi_k = \pi_k(1-\pi_k)\bm{x}_k$, the gradient of the loss in \eqref{eq:logisticv1} is:
\begin{equation*}
\nabla_{\bm{\theta}} \sum_{i,j=1}^n K_x(\bm{x}_i, \bm{x}_j) l(\bm{\theta}, \bm{x}_i, \bm{x}_j, y_i) = \sum_{i,j=1}^n K_x(\bm{x}_i, \bm{x}_j) \left( \nabla_{\bm{\theta}} \text{Term 1} - \nabla_{\bm{\theta}} \text{Term 2} \right),
\end{equation*}
where:
\begin{align*}
\nabla_{\bm{\theta}} \text{Term 1} &= 0.5h_y^2 \left[ \pi_i(1-\pi_i)(2\pi_j - 1)\bm{x}_i + \pi_j(1-\pi_j)(2\pi_i - 1)\bm{x}_j \right], \\
\nabla_{\bm{\theta}} \text{Term 2} &= h_y \pi_j(1-\pi_j) \left[ (1-h_y)^{1-y_i} - (1-h_y)^{y_i} \right] \bm{x}_j.
\end{align*}

\section{Simulation Studies}\label{sec:num}

In this section we carry out two comprehensive simulation studies to compare the proposed MMD-based estimators $\hat{\bm \theta}_n$ and $\tilde{\bm \theta}_n$ against some benchmark methods under the Gaussian linear regression and logistic regression settings. 

Some tuning parameters are chosen in the same way in all simulations. 
For the optimization of our proposed estimators, we set the ADMM penalty parameter to $\rho = 1$, a standard default that provides a robust balance between primal and dual convergence in consensus optimization \cite{boyd2011distributed}. 
Within the ADMM steps, the $\bm \theta$-update is performed using the AdaGrad algorithm with a learning rate of 0.1, which is widely used for its ability to adaptively scale gradients in non-convex landscapes \cite{duchi2011adaptive}. 
The maximum iterations (500 inner, 200 outer) and tolerances ($10^{-4}, 10^{-3}$) follow standard practices in high-dimensional regularized regression to ensure numerical stability without excessive computational cost.

A critical component of MMD-based estimation is the choice of bandwidth parameters for the kernels $K_x$ and $K_y$. For each replicate, we utilize the \textit{median heuristic} to adaptively set $h_y$ and $h_x$. Specifically, $h_y$ is set to the median of the pairwise Euclidean distances of the response variables $\{y_i\}_{i=1}^n$, and $h_x$ is set to the median of the pairwise distances of the input vectors $\{\bm{x}_i\}_{i=1}^n$. 
However, for the binary response, the median heuristic is not applicable due to the discrete nature of the labels. 
We fix $h_y = \sqrt{2}/2 \approx 0.707$. 
In the context of the geometric kernel for discrete data, this choice ensures that the kernel assigns a significantly higher weight to matching labels ($y_i = y_j$) compared to mismatches, while maintaining enough smoothing to prevent the MMD distance from becoming overly sensitive to individual label flips. 
This value has been found to provide stable performance across a wide range of classification tasks in MMD-related literature \cite{alquier2024universal,rajagopalan1995kernel}.

\subsection{Gaussian Linear Regression}

To evaluate the performance of our proposed penalized MMD estimators, we first consider a high-dimensional linear regression model:
\begin{equation}
y_i = \bm{x}_i^\top \bm{\beta} + \epsilon_i, \quad i = 1, \dots, n,
\end{equation}
where we set $n=100$ and $p=200$. The true parameter vector is defined as a sparse vector $\bm{\beta} = (4, 4, 3, 3, -3, -3, -4, -4, \bm{0}_{p-8})^\top$. 
The input variables $\bm{x}_i$ are generated from a multivariate normal distribution $\mathcal{N}(\bm{0}, \bm{\Sigma}_X)$. We consider two covariance structures for $\bm{\Sigma}_X$: (i) an identity matrix $\bm{I}_p$ and (ii) an autoregressive (AR) structure where $(\bm{\Sigma}_X)_{jk} = 0.7^{|j-k|}$.

To test the ``universal'' robustness of the estimators, we vary the distribution of the error term $\epsilon_i$ across three settings: standard normal $\mathcal{N}(0, 1)$, Laplace$(0, 1)$, and a heavy-tailed Student's $t$-distribution with 5 degrees of freedom ($t_5$).
We introduce three levels of contamination $\tau \in \{0, 0.05, 0.1\}$ and three distinct outlier types $\mathcal{T}$:
(1) $\mathcal{X}_1$: A single predictor $x_{i,2}$ is randomly replaced by values from $\mathcal{N}(5, 1)$, creating leverage points.
(2) $\mathcal{X}_2$: Two predictors $x_{i,2}$ and $x_{i,5}$ are replaced by $\mathcal{N}(5, 1)$, creating high-leverage contamination.
(3) $\mathcal{Y}$: The response variable $y_i$ is replaced by $\mathcal{N}(10, 1)$, representing standard response outliers.

We compare our proposed $O(n^2)$ estimator $\hat{\bm{\theta}}_n$ and $O(n)$ reduced estimator $\tilde{\bm{\theta}}_n$ with three benchmarks: the standard Lasso, the Huber regression with $\ell_1$ penalty, and Loh's robust $M$-estimator \citep{loh2017statistical}. 
%For the MMD estimators, we utilize the median heuristic to adaptively set the bandwidth parameters $h_y$ and $h_x$. 
All methods use 5-fold cross-validation to select the optimal tuning parameter $\lambda$.

We evaluate the performance of all methods using three measures: $\text{MSE}=||\bm \beta-\hat{\bm \beta}||$, where $\hat{\bm \beta}$ is the returned estimation by any method; $FSL=FP+FN$ represents the counts of false positive (FP) and false negative (FN) in variable selection; $\text{PE}=\frac{1}{100}||\bm y_{\text{test}}-\bm X_{\text{test}}\hat{\bm \beta}||^2$ is the mean square prediction error for a separate generated test data set of 100 samples without any outliers or contamination. 
The simulation results over 50 replicates are summarized in Table~\ref{tab:mse_results} (MSE), Table~\ref{tab:fsl_results} (FSL), and Table~\ref{tab:pe_results} (PE). 

{\bf Estimation Accuracy (MSE):} As shown in Table~\ref{tab:mse_results}, the Huber estimator is highly effective when the contamination is limited to the response variable ($\mathcal{Y}$) under normal errors. However, our proposed MMD estimators $\hat{\bm \theta}_n$ and $\tilde{\bm \theta}_n$ demonstrate superior robustness when moving to heavy-tailed error distributions ($Laplace$ and $t_5$). Notably, Lasso and Loh's estimator show a significant increase in MSE under $\mathcal{X}_2$ contamination, while the proposed MMD framework remains stable, particularly in the $\bm \Sigma_{X}=$AR(0.7) setting.

{\bf Variable Selection (FSL):} Table~\ref{tab:fsl_results} highlights a primary strength of the penalized MMD approach. Standard Lasso and Huber regression tend to select a large number of irrelevant features (high FSL) when outliers are present. In contrast, $\hat{\bm \theta}_n$ and $\tilde{\bm \theta}_n$ consistently achieve much lower FSL values, often comparable to or better than Loh's estimator. This indicates that the MMD loss function is less likely to be ``distracted'' by contaminated samples during the feature selection process.

{\bf Predictive Performance (PE):} The prediction error results in Table~\ref{tab:pe_results} confirm that the robustness of the MMD estimators translates to better generalization. While Huber performs well for $Y$-outliers, its performance degrades more than the MMD methods under $X$-leverage points. Furthermore, the $O(n)$ approximate estimator $\tilde{\bm \theta}_n$ performs remarkably close to the full $O(n^2)$ version $\hat{\bm \theta}_n$ across all settings, providing empirical justification for its use in larger datasets to save computational cost without sacrificing significant accuracy.

Overall, the MMD-based estimators prove to be the most versatile. 
The Huber estimator remains the most effective specifically for response-variable outliers ($\mathcal{Y}$) under normal errors. 
The proposed MMD estimators provide a more consistent performance across the full spectrum of outlier types. 
Specifically, as shown in Table \ref{tab:fsl_results}, the MMD framework consistently achieves lower FSL than the Huber and Lasso benchmarks, suggesting that it is less likely to be misled by outliers during the feature selection process. This 'universal' stability is a key advantage of the MMD-based joint-distribution perspective.
While Huber may lead in specific response-only outlier scenarios, the MMD methods demonstrate a ``universal'' robustness that covers leverage points, response outliers, and non-normal errors simultaneously. The results also suggest that the proposed methods are particularly effective when predictors are correlated (AR structure), which is common in high-dimensional real-world data.

\begin{table}[htbp]
\centering
\caption{The averages and standard errors (in parenthesis) of MSE for estimates. The smallest mean in each row is in bold.}
\label{tab:mse_results}
\footnotesize
\begin{tabular}{cccccccc}
\toprule
Error Dist. & $\tau$ & $\mathcal{T}$ & Lasso & Huber & Loh & $\hat{\bm \theta}_n$ & $\tilde{\bm \theta}_n$ \\
\midrule
\multicolumn{8}{c}{\textbf{Panel A: Identity Matrix $\bm{\Sigma}_X = \bm{I}_p$}} \\
\midrule
\multirow{7}{*}{$N(0,1)$} & 0 & -- & 0.00220 (0.0011) & 0.00258 (0.0013) & 0.00230 (0.0012) & 0.00207 (0.0011) & \textbf{0.00204} (0.0010) \\
\cmidrule(lr){2-8}
& \multirow{3}{*}{0.05} & $\mathcal{X}_1$ & 0.0729 (0.021) & \textbf{0.0397} (0.022) & 0.0662 (0.014) & 0.0725 (0.020) & 0.0722 (0.018) \\
& & $\mathcal{X}_2$ & 0.0159 (0.0099) & \textbf{0.00797} (0.0047) & 0.0107 (0.0066) & 0.0161 (0.011) & 0.0166 (0.012) \\
& & $\mathcal{Y}$ & 0.0272 (0.015) & \textbf{0.00480} (0.0019) & 0.0246 (0.014) & 0.0268 (0.015) & 0.0249 (0.014) \\
\cmidrule(lr){2-8}
& \multirow{3}{*}{0.1} & $\mathcal{X}_1$ & 0.0847 (0.012) & 0.0815 (0.020) & \textbf{0.0783} (0.016) & 0.0854 (0.011) & 0.0823 (0.011) \\
& & $\mathcal{X}_2$ & 0.0262 (0.012) & 0.0121 (0.0091) & \textbf{0.0116} (0.0046) & 0.0258 (0.013) & 0.0254 (0.011) \\
& & $\mathcal{Y}$ & 0.0433 (0.015) & \textbf{0.00856} (0.0042) & 0.0419 (0.016) & 0.0420 (0.015) & 0.0396 (0.015) \\
\midrule
\multirow{7}{*}{Laplace} & 0 & -- & 0.00549 (0.00081) & 0.00530 (0.0020) & 0.00625 (0.0017) & 0.00494 (0.0011) & \textbf{0.00482} (0.0011) \\
\cmidrule(lr){2-8}
& \multirow{3}{*}{0.05} & $\mathcal{X}_1$ & 0.0670 (0.013) & \textbf{0.0392} (0.021) & 0.0636 (0.021) & 0.0673 (0.014) & 0.0679 (0.014) \\
& & $\mathcal{X}_2$ & 0.0162 (0.0052) & \textbf{0.0104} (0.0059) & 0.0127 (0.0047) & 0.0164 (0.0060) & 0.0165 (0.0059) \\
& & $\mathcal{Y}$ & 0.0302 (0.011) & \textbf{0.00665} (0.0027) & 0.0302 (0.010) & 0.0340 (0.016) & 0.0262 (0.0081) \\
\cmidrule(lr){2-8}
& \multirow{3}{*}{0.1} & $\mathcal{X}_1$ & 0.0782 (0.015) & 0.0820 (0.018) & \textbf{0.0756} (0.020) & 0.0781 (0.014) & 0.0778 (0.016) \\
& & $\mathcal{X}_2$ & 0.0393 (0.027) & \textbf{0.0205} (0.014) & 0.0224 (0.015) & 0.0386 (0.028) & 0.0354 (0.022) \\
& & $\mathcal{Y}$ & 0.0625 (0.040) & \textbf{0.0149} (0.0087) & 0.0636 (0.041) & 0.0628 (0.038) & 0.0642 (0.050) \\
\midrule
\multirow{7}{*}{$t_5$} & 0 & -- & 0.00406 (0.0013) & 0.00408 (0.0014) & \textbf{0.00378} (0.00084) & 0.00382 (0.0013) & 0.00381 (0.0012) \\
\cmidrule(lr){2-8}
& \multirow{3}{*}{0.05} & $\mathcal{X}_1$ & 0.0813 (0.015) & \textbf{0.0382} (0.016) & 0.0728 (0.026) & 0.0818 (0.015) & 0.0771 (0.014) \\
& & $\mathcal{X}_2$ & 0.0174 (0.014) & \textbf{0.00824} (0.0048) & 0.0112 (0.0083) & 0.0168 (0.013) & 0.0176 (0.016) \\
& & $\mathcal{Y}$ & 0.0215 (0.0072) & \textbf{0.00652} (0.0027) & 0.0220 (0.0083) & 0.0205 (0.0066) & 0.0205 (0.0059) \\
\cmidrule(lr){2-8}
& \multirow{3}{*}{0.1} & $\mathcal{X}_1$ & 0.0843 (0.016) & 0.0898 (0.024) & \textbf{0.0802} (0.019) & 0.0855 (0.016) & 0.0855 (0.020) \\
& & $\mathcal{X}_2$ & 0.0281 (0.015) & 0.0141 (0.0076) & \textbf{0.0131} (0.0068) & 0.0269 (0.015) & 0.0259 (0.013) \\
& & $\mathcal{Y}$ & 0.0572 (0.024) & \textbf{0.0111} (0.0040) & 0.0557 (0.023) & 0.0568 (0.023) & 0.0557 (0.024) \\
\midrule
\multicolumn{8}{c}{\textbf{Panel B: AR(0.7) Matrix $\bm{\Sigma}_X = \text{AR}(0.7)$}} \\
\midrule
\multirow{7}{*}{$N(0,1)$} & 0 & -- & 0.00393 (0.0014) & 0.00410 (0.0015) & 0.00469 (0.0019) & \textbf{0.00379} (0.0014) & 0.00389 (0.0013) \\
\cmidrule(lr){2-8}
& \multirow{3}{*}{0.05} & $\mathcal{X}_1$ & 0.114 (0.029) & \textbf{0.0920} (0.038) & 0.0983 (0.029) & 0.113 (0.025) & 0.112 (0.025) \\
& & $\mathcal{X}_2$ & 0.0323 (0.020) & \textbf{0.0139} (0.0090) & 0.0156 (0.0083) & 0.0325 (0.018) & 0.0340 (0.020) \\
& & $\mathcal{Y}$ & 0.0349 (0.018) & \textbf{0.00641} (0.0031) & 0.0404 (0.019) & 0.0355 (0.018) & 0.0324 (0.016) \\
\cmidrule(lr){2-8}
& \multirow{3}{*}{0.1} & $\mathcal{X}_1$ & 0.125 (0.030) & 0.126 (0.034) & \textbf{0.117} (0.018) & 0.122 (0.027) & 0.123 (0.027) \\
& & $\mathcal{X}_2$ & 0.0700 (0.029) & 0.0383 (0.020) & \textbf{0.0295} (0.014) & 0.0688 (0.023) & 0.0670 (0.021) \\
& & $\mathcal{Y}$ & 0.0906 (0.035) & \textbf{0.0152} (0.0092) & 0.0805 (0.033) & 0.0928 (0.038) & 0.0911 (0.032) \\
\midrule
\multirow{7}{*}{Laplace} & 0 & -- & 0.00903 (0.0033) & 0.00891 (0.0040) & 0.00836 (0.0038) & 0.00840 (0.0037) & \textbf{0.00834} (0.0034) \\
\cmidrule(lr){2-8}
& \multirow{3}{*}{0.05} & $\mathcal{X}_1$ & 0.0812 (0.017) & \textbf{0.0647} (0.037) & 0.0703 (0.013) & 0.0834 (0.015) & 0.0826 (0.019) \\
& & $\mathcal{X}_2$ & 0.0168 (0.0079) & 0.0125 (0.0057) & \textbf{0.0107} (0.0036) & 0.0160 (0.0062) & 0.0156 (0.0071) \\
& & $\mathcal{Y}$ & 0.0605 (0.027) & \textbf{0.0168} (0.0062) & 0.0553 (0.030) & 0.0615 (0.030) & 0.0539 (0.026) \\
\cmidrule(lr){2-8}
& \multirow{3}{*}{0.1} & $\mathcal{X}_1$ & 0.119 (0.022) & 0.129 (0.032) & \textbf{0.0998} (0.020) & 0.121 (0.020) & 0.121 (0.022) \\
& & $\mathcal{X}_2$ & 0.0581 (0.028) & 0.0427 (0.031) & \textbf{0.0292} (0.016) & 0.0582 (0.026) & 0.0575 (0.030) \\
& & $\mathcal{Y}$ & 0.0830 (0.025) & \textbf{0.0175} (0.0026) & 0.0804 (0.034) & 0.0809 (0.025) & 0.0852 (0.031) \\
\midrule
\multirow{7}{*}{$t_5$} & 0 & -- & 0.00655 (0.0017) & 0.00621 (0.0019) & 0.00612 (0.0017) & \textbf{0.00607} (0.0017) & 0.00664 (0.0020) \\
\cmidrule(lr){2-8}
& \multirow{3}{*}{0.05} & $\mathcal{X}_1$ & 0.100 (0.027) & 0.0945 (0.045) & \textbf{0.0912} (0.025) & 0.105 (0.030) & 0.100 (0.028) \\
& & $\mathcal{X}_2$ & 0.0400 (0.039) & \textbf{0.0169} (0.0040) & 0.0215 (0.015) & 0.0389 (0.034) & 0.0368 (0.030) \\
& & $\mathcal{Y}$ & 0.0443 (0.034) & \textbf{0.00876} (0.0047) & 0.0410 (0.038) & 0.0439 (0.033) & 0.0386 (0.029) \\
\cmidrule(lr){2-8}
& \multirow{3}{*}{0.1} & $\mathcal{X}_1$ & 0.130 (0.035) & 0.141 (0.037) & \textbf{0.113} (0.029) & 0.132 (0.036) & 0.129 (0.034) \\
& & $\mathcal{X}_2$ & 0.0729 (0.048) & 0.0428 (0.033) & \textbf{0.0280} (0.016) & 0.0708 (0.047) & 0.0690 (0.043) \\
& & $\mathcal{Y}$ & 0.0984 (0.033) & \textbf{0.0175} (0.011) & 0.0911 (0.035) & 0.0957 (0.036) & 0.0900 (0.033) \\
\bottomrule
\end{tabular}
\end{table}

\begin{table}[htbp]
\centering
\caption{The averages and standard errors (in parenthesis) of FSL for estimates. The smallest mean in each row is in bold.}
\label{tab:fsl_results}
\footnotesize
\begin{tabular}{cccccccc}
\toprule
Error Dist. & $\tau$ & $\mathcal{T}$ & Lasso & Huber & Loh & $\hat{\bm \theta}_n$ & $\tilde{\bm \theta}_n$\\
\midrule
\multicolumn{8}{c}{\textbf{Panel A: Identity Matrix $\bm{\Sigma}_X = \bm{I}_p$}} \\
\midrule
\multirow{7}{*}{$N(0,1)$} & 0 & -- & 20.6 (7.9) & 24.1 (9.1) & 4.30 (2.0) & 3.30 (3.7) & \textbf{3.00} (2.1) \\
\cmidrule(lr){2-8}
& \multirow{3}{*}{0.05} & $\mathcal{X}_1$ & 27.0 (12) & 26.7 (9.4) & 20.6 (10) & \textbf{15.8} (8.1) & 17.8 (14) \\
& & $\mathcal{X}_2$ & 27.6 (9.5) & 29.2 (7.5) & 12.4 (6.5) & 9.70 (5.0) & \textbf{3.80} (5.5) \\
& & $\mathcal{Y}$ & 18.0 (6.6) & 19.0 (5.8) & 14.0 (5.3) & 11.3 (6.8) & \textbf{4.40} (3.5) \\
\cmidrule(lr){2-8}
& \multirow{3}{*}{0.1} & $\mathcal{X}_1$ & 23.7 (9.2) & 22.5 (9.6) & 19.4 (8.2) & 14.7 (7.1) & \textbf{11.9} (5.1) \\
& & $\mathcal{X}_2$ & 26.1 (10) & 27.2 (9.3) & \textbf{8.30} (4.8) & 13.4 (5.9) & 9.70 (10) \\
& & $\mathcal{Y}$ & 26.4 (15) & 16.8 (5.8) & 23.3 (10) & \textbf{13.1} (7.7) & 15.4 (14) \\
\midrule
\multirow{7}{*}{Laplace} & 0 & -- & 32.5 (11) & 29.6 (17) & 13.4 (4.9) & 9.40 (5.2) & \textbf{7.60} (2.4) \\
\cmidrule(lr){2-8}
& \multirow{3}{*}{0.05} & $\mathcal{X}_1$ & 27.5 (13) & 26.0 (7.6) & 23.6 (7.0) & \textbf{19.0} (13) & 22.1 (19) \\
& & $\mathcal{X}_2$ & 33.8 (11) & 26.6 (12) & 14.7 (7.7) & 16.2 (9.2) & \textbf{13.8} (8.8) \\
& & $\mathcal{Y}$ & 25.1 (8.9) & 16.1 (4.6) & 21.8 (5.4) & 10.5 (7.5) & \textbf{8.50} (4.0) \\
\cmidrule(lr){2-8}
& \multirow{3}{*}{0.1} & $\mathcal{X}_1$ & 35.5 (17) & 29.8 (12) & 25.7 (12) & 23.9 (14) & \textbf{21.7} (12) \\
& & $\mathcal{X}_2$ & 36.7 (10) & 34.4 (6.9) & \textbf{17.6} (7.3) & 20.6 (7.3) & 19.7 (17) \\
& & $\mathcal{Y}$ & 26.3 (13) & 19.4 (6.3) & 24.3 (9.5) & \textbf{15.7} (9.0) & 18.2 (15) \\
\midrule
\multirow{7}{*}{$t_5$} & 0 & -- & 24.9 (10) & 28.2 (12) & 7.80 (2.3) & 6.50 (4.6) & \textbf{4.50} (2.7) \\
\cmidrule(lr){2-8}
& \multirow{3}{*}{0.05} & $\mathcal{X}_1$ & 25.2 (14) & 30.3 (14) & 11.0 (5.8) & 10.2 (6.3) & \textbf{8.40} (6.4) \\
& & $\mathcal{X}_2$ & 25.2 (14) & 30.3 (14) & 11.0 (5.8) & 10.2 (6.3) & \textbf{8.40} (6.4) \\
& & $\mathcal{Y}$ & 21.4 (9.2) & 15.3 (5.8) & 19.4 (8.6) & \textbf{12.0} (5.3) & 12.1 (9.9) \\
\cmidrule(lr){2-8}
& \multirow{3}{*}{0.1} & $\mathcal{X}_1$ & 23.7 (9.2) & 25.0 (10) & 20.7 (7.7) & \textbf{15.2} (7.6) & 17.2 (20) \\
& & $\mathcal{X}_2$ & 38.2 (15) & 33.9 (11) & \textbf{11.7} (9.5) & 20.4 (12) & 14.1 (14) \\
& & $\mathcal{Y}$ & 22.1 (9.8) & 22.3 (6.8) & 19.4 (4.6) & 17.6 (6.1) & \textbf{14.2} (13) \\
\midrule
\multicolumn{8}{c}{\textbf{Panel B: AR(0.7) Matrix $\bm{\Sigma}_X = \text{AR}(0.7)$}} \\
\midrule
\multirow{7}{*}{$N(0,1)$} & 0 & -- & 20.2 (9.3) & 21.9 (11) & 3.50 (1.4) & \textbf{3.30} (3.0) & 3.80 (3.6) \\
\cmidrule(lr){2-8}
& \multirow{3}{*}{0.05} & $\mathcal{X}_1$ & 22.5 (14) & 25.3 (11) & 14.1 (6.1) & 17.0 (12) & \textbf{12.0} (10) \\
& & $\mathcal{X}_2$ & 32.4 (15) & 26.9 (8.0) & \textbf{9.20} (5.7) & 12.0 (6.9) & 9.30 (7.5) \\
& & $\mathcal{Y}$ & 23.5 (17) & 20.4 (11) & 17.2 (8.0) & 9.90 (7.1) & \textbf{8.90} (7.1) \\
\cmidrule(lr){2-8}
& \multirow{3}{*}{0.1} & $\mathcal{X}_1$ & 18.1 (7.1) & 21.6 (6.4) & 18.7 (10) & 13.9 (5.8) & \textbf{10.3} (8.1) \\
& & $\mathcal{X}_2$ & 32.7 (15) & 35.2 (12) & \textbf{10.8} (5.4) & 16.9 (8.7) & 14.9 (13) \\
& & $\mathcal{Y}$ & 19.0 (12) & 19.0 (6.1) & 18.2 (6.3) & 12.4 (9.4) & \textbf{8.40} (6.0) \\
\midrule
\multirow{7}{*}{Laplace} & 0 & -- & 19.1 (7.6) & 14.2 (5.0) & 9.70 (4.0) & 7.90 (3.0) & \textbf{4.80} (2.4) \\
\cmidrule(lr){2-8}
& \multirow{3}{*}{0.05} & $\mathcal{X}_1$ & 21.7 (5.7) & 22.6 (7.7) & 17.6 (9.5) & \textbf{10.1} (6.1) & 10.8 (7.9) \\
& & $\mathcal{X}_2$ & 25.1 (9.6) & 24.2 (9.4) & 11.8 (5.7) & 12.4 (6.2) & \textbf{7.30} (4.9) \\
& & $\mathcal{Y}$ & 18.2 (15) & 18.1 (6.4) & 17.8 (6.9) & 13.4 (13) & \textbf{9.80} (11) \\
\cmidrule(lr){2-8}
& \multirow{3}{*}{0.1} & $\mathcal{X}_1$ & 22.2 (10) & 29.0 (15) & 17.0 (9.2) & \textbf{11.1} (7.4) & 15.1 (9.9) \\
& & $\mathcal{X}_2$ & 33.7 (12) & 31.6 (9.5) & 14.0 (5.8) & 18.0 (8.4) & \textbf{11.9} (8.3) \\
& & $\mathcal{Y}$ & 14.7 (8.3) & 20.7 (11) & 17.7 (13) & 9.30 (6.5) & \textbf{8.80} (8.0) \\
\midrule
\multirow{7}{*}{$t_5$} & 0 & -- & 20.3 (6.9) & 21.8 (12) & 5.70 (2.7) & 5.90 (2.0) & \textbf{3.60} (2.0) \\
\cmidrule(lr){2-8}
& \multirow{3}{*}{0.05} & $\mathcal{X}_1$ & 25.0 (4.1) & 21.6 (7.1) & 16.6 (7.8) & 14.1 (7.7) & \textbf{11.8} (6.6) \\
& & $\mathcal{X}_2$ & 28.6 (9.0) & 25.5 (7.9) & 13.8 (4.5) & 14.0 (5.5) & \textbf{10.7} (5.3) \\
& & $\mathcal{Y}$ & 20.5 (8.7) & 18.8 (5.4) & 16.4 (3.1) & 10.5 (7.7) & \textbf{8.00} (7.8) \\
\cmidrule(lr){2-8}
& \multirow{3}{*}{0.1} & $\mathcal{X}_1$ & 22.0 (8.6) & 23.7 (13) & 16.7 (4.7) & 14.1 (6.1) & \textbf{12.9} (7.6) \\
& & $\mathcal{X}_2$ & 29.4 (11) & 33.3 (8.8) & \textbf{10.6} (6.6) & 14.3 (6.3) & 12.7 (13) \\
& & $\mathcal{Y}$ & 15.6 (8.7) & 14.9 (4.5) & 18.5 (4.6) & 11.1 (6.8) & \textbf{7.70} (4.3) \\
\bottomrule
\end{tabular}
\end{table}

\begin{table}[htbp]
\centering
\caption{The averages and standard errors (in parenthesis) of PE for estimates. The smallest mean in each row is in bold.}
\label{tab:pe_results}
\footnotesize
\begin{tabular}{cccccccc}
\toprule
Error Dist. & $\tau$ & $\mathcal{T}$ & Lasso & Huber & Loh & $\hat{\bm \theta}_n$ & $\tilde{\bm \theta}_n$ \\
\midrule
\multicolumn{8}{c}{\textbf{Panel A: Identity Matrix $\bm{\Sigma}_X = \bm{I}_p$}} \\
\midrule
\multirow{7}{*}{$N(0,1)$} & 0 & -- & 1.42 (0.32) & 1.49 (0.31) & 1.47 (0.37) & 1.41 (0.34) & \textbf{1.39} (0.31) \\
\cmidrule(lr){2-8}
& \multirow{3}{*}{0.05} & $\mathcal{X}_1$ & 17.1 (6.3) & \textbf{10.1} (5.9) & 15.1 (3.5) & 17.1 (5.6) & 17.0 (5.7) \\
& & $\mathcal{X}_2$ & 4.07 (2.0) & \textbf{2.59} (0.90) & 3.14 (1.4) & 4.12 (2.3) & 4.24 (2.4) \\
& & $\mathcal{Y}$ & 7.02 (3.7) & \textbf{2.09} (0.64) & 6.35 (3.1) & 6.93 (3.7) & 6.51 (3.3) \\
\cmidrule(lr){2-8}
& \multirow{3}{*}{0.1} & $\mathcal{X}_1$ & 17.7 (3.7) & 17.3 (3.8) & \textbf{16.8} (4.7) & 17.9 (3.4) & 17.2 (3.4) \\
& & $\mathcal{X}_2$ & 6.71 (3.0) & 3.69 (2.2) & \textbf{3.55} (1.3) & 6.62 (3.1) & 6.51 (2.8) \\
& & $\mathcal{Y}$ & 9.88 (3.0) & \textbf{2.83} (0.90) & 9.53 (3.7) & 9.53 (3.0) & 9.25 (2.7) \\
\midrule
\multirow{7}{*}{Laplace} & 0 & -- & 3.18 (0.56) & 3.16 (0.72) & 3.32 (0.74) & 3.10 (0.65) & \textbf{3.07} (0.67) \\
\cmidrule(lr){2-8}
& \multirow{3}{*}{0.05} & $\mathcal{X}_1$ & 15.4 (2.2) & \textbf{10.1} (5.1) & 14.9 (3.9) & 15.7 (2.5) & 15.4 (2.5) \\
& & $\mathcal{X}_2$ & 4.99 (0.79) & \textbf{3.99} (1.4) & 4.29 (0.77) & 5.13 (1.1) & 5.06 (0.83) \\
& & $\mathcal{Y}$ & 7.23 (2.2) & \textbf{3.35} (0.78) & 7.32 (2.1) & 7.86 (2.9) & 6.76 (2.0) \\
\cmidrule(lr){2-8}
& \multirow{3}{*}{0.1} & $\mathcal{X}_1$ & 17.2 (2.2) & 19.6 (4.3) & \textbf{16.5} (3.5) & 17.2 (2.1) & 16.5 (2.3) \\
& & $\mathcal{X}_2$ & 9.22 (5.0) & \textbf{5.82} (2.9) & 6.18 (3.2) & 9.05 (5.2) & 8.67 (4.6) \\
& & $\mathcal{Y}$ & 15.1 (8.8) & \textbf{5.02} (1.6) & 15.0 (8.6) & 15.2 (8.3) & 15.5 (11) \\
\midrule
\multirow{7}{*}{$t_5$} & 0 & -- & 2.31 (0.35) & 2.28 (0.35) & \textbf{2.21} (0.23) & 2.28 (0.32) & 2.27 (0.29) \\
\cmidrule(lr){2-8}
& \multirow{3}{*}{0.05} & $\mathcal{X}_1$ & 17.2 (2.2) & 18.8 (6.3) & 17.4 (3.5) & 18.3 (3.9) & \textbf{16.5} (2.3) \\
& & $\mathcal{X}_2$ & 4.91 (3.2) & \textbf{3.02} (1.2) & 3.77 (2.0) & 4.83 (3.1) & 5.14 (4.0) \\
& & $\mathcal{Y}$ & 5.60 (1.5) & \textbf{2.80} (0.74) & 5.78 (2.1) & 5.53 (1.5) & 5.45 (1.5) \\
\cmidrule(lr){2-8}
& \multirow{3}{*}{0.1} & $\mathcal{X}_1$ & 17.9 (3.5) & 18.8 (6.3) & 17.4 (3.5) & 18.3 (3.9) & \textbf{16.5} (2.3) \\
& & $\mathcal{X}_2$ & 4.91 (3.2) & \textbf{3.02} (1.2) & 3.77 (2.0) & 4.83 (3.1) & 5.14 (4.0) \\
& & $\mathcal{Y}$ & 13.2 (4.6) & \textbf{3.90} (0.96) & 12.9 (4.7) & 13.1 (4.4) & 13.2 (5.2) \\
\midrule
\multicolumn{8}{c}{\textbf{Panel B: AR(0.7) Matrix $\bm{\Sigma}_X = \text{AR}(0.7)$}} \\
\midrule
\multirow{7}{*}{$N(0,1)$} & 0 & -- & 1.32 (0.24) & 1.34 (0.21) & 1.32 (0.20) & \textbf{1.27} (0.21) & 1.29 (0.21) \\
\cmidrule(lr){2-8}
& \multirow{3}{*}{0.05} & $\mathcal{X}_1$ & 11.4 (2.0) & \textbf{9.30} (3.5) & 10.4 (2.4) & 11.8 (2.8) & 11.5 (2.8) \\
& & $\mathcal{X}_2$ & 3.69 (1.4) & \textbf{2.24} (0.74) & 2.35 (0.53) & 3.53 (1.2) & 3.65 (1.3) \\
& & $\mathcal{Y}$ & 5.81 (2.9) & \textbf{1.77} (0.38) & 7.00 (3.6) & 5.97 (3.3) & 5.34 (2.7) \\
\cmidrule(lr){2-8}
& \multirow{3}{*}{0.1} & $\mathcal{X}_1$ & 11.3 (1.5) & 11.4 (1.7) & 11.4 (2.1) & \textbf{11.1} (1.2) & 11.7 (1.4) \\
& & $\mathcal{X}_2$ & 6.66 (2.6) & 4.27 (1.8) & \textbf{3.71} (0.92) & 6.50 (2.2) & 6.83 (2.5) \\
& & $\mathcal{Y}$ & 12.8 (6.3) & \textbf{2.43} (0.42) & 10.9 (4.6) & 13.0 (6.1) & 11.9 (4.9) \\
\midrule
\multirow{7}{*}{Laplace} & 0 & -- & 2.95 (0.07) & \textbf{2.80} (0.07) & 3.06 (0.08) & 2.99 (0.09) & 2.99 (0.09) \\
\cmidrule(lr){2-8}
& \multirow{3}{*}{0.05} & $\mathcal{X}_1$ & 10.3 (0.33) & \textbf{8.59} (0.41) & 9.54 (0.30) & 9.49 (0.65) & 9.49 (0.65) \\
& & $\mathcal{X}_2$ & 5.02 (0.21) & \textbf{3.79} (0.13) & 3.99 (0.16) & 4.05 (0.17) & 4.05 (0.17) \\
& & $\mathcal{Y}$ & 8.74 (0.61) & \textbf{3.29} (0.14) & 8.26 (0.51) & 5.59 (0.40) & 5.59 (0.40) \\
\cmidrule(lr){2-8}
& \multirow{3}{*}{0.1} & $\mathcal{X}_1$ & 12.2 (0.36) & 13.0 (0.43) & \textbf{11.6} (0.40) & 13.1 (0.51) & 13.1 (0.51) \\
& & $\mathcal{X}_2$ & 6.80 (0.41) & 5.24 (0.33) & \textbf{4.57} (0.19) & 5.15 (0.34) & 5.15 (0.34) \\
& & $\mathcal{Y}$ & 15.1 (0.97) & \textbf{4.68} (0.25) & 13.9 (0.79) & 11.0 (0.76) & 11.0 (0.76) \\
\midrule
\multirow{7}{*}{$t_5$} & 0 & -- & 2.27 (0.08) & \textbf{2.22} (0.07) & 2.34 (0.08) & 2.47 (0.10) & 2.47 (0.10) \\
\cmidrule(lr){2-8}
& \multirow{3}{*}{0.05} & $\mathcal{X}_1$ & 9.90 (0.37) & \textbf{8.45} (0.50) & 9.28 (0.38) & 9.14 (0.54) & 9.14 (0.54) \\
& & $\mathcal{X}_2$ & 4.17 (0.21) & 3.14 (0.15) & 3.26 (0.14) & 3.23 (0.18) & \textbf{3.14} (0.15) \\
& & $\mathcal{Y}$ & 6.88 (0.56) & \textbf{2.66} (0.09) & 7.07 (0.66) & 4.26 (0.33) & 4.26 (0.33) \\
\cmidrule(lr){2-8}
& \multirow{3}{*}{0.1} & $\mathcal{X}_1$ & 11.5 (0.33) & 12.3 (0.42) & \textbf{11.1} (0.35) & 11.4 (0.45) & 11.4 (0.45) \\
& & $\mathcal{X}_2$ & 6.75 (0.29) & 5.08 (0.25) & \textbf{4.38} (0.21) & 5.00 (0.35) & 5.00 (0.35) \\
& & $\mathcal{Y}$ & 13.7 (0.85) & \textbf{3.26} (0.14) & 13.5 (0.86) & 9.98 (0.75) & 9.98 (0.75) \\
\bottomrule
\end{tabular}
\end{table}

\subsection{Binary Logistic Regression}

We further evaluate the performance of the proposed estimators in the context of high-dimensional binary classification. The response variables $y_i \in \{0, 1\}$ are generated from a Bernoulli distribution with probability $p(y_i=1) = [1 + \exp(-\bm{x}_i^\top \bm{\beta})]^{-1}$. 
We consider two problem sizes: $(n=100, p=100)$ and $(n=100, p=200)$. 
The true parameter vector is sparse, with $\bm{\beta} = (1, \dots, 1, \bm{0}_{p-10})^\top$, where the first ten elements are set to one. 
The input variables $\bm{x}_i$ are generated from $\mathcal{N}(\bm{0}, \bm{\Sigma}_X)$ using both the identity and AR(0.7) covariance structures. 

To assess robustness, we introduce contamination $\tau \in \{0, 0.05, 0.1\}$ across four distinct outlier types $\mathcal{T}$: 
(1) $\mathcal{X}_1$: Predictors $x_{i,2}$ and $x_{i,5}$ are replaced by values from $\mathcal{N}(20, 1)$.
(2) $\mathcal{X}_2$: All active predictors $x_{i,1}, \dots, x_{i,10}$ are replaced by $\mathcal{N}(20, 1)$.
(3) $\mathcal{XY}_1$: Labels $y_i$ are flipped to $1-y_i$ and their corresponding active predictors are replaced by $\mathcal{N}(20, 1)$.
(4) $\mathcal{XY}_2$: A more extreme case where the label flipping and predictor replacement occur on different subsets of the contaminated samples.

The proposed full estimator $\hat{\bm{\theta}}$ and reduced estimator $\tilde{\bm{\theta}}$ are compared against the standard logistic Lasso and the robust elastic net least trimmed squares (enetLTS) estimator \citep{kurnaz2018robust} implemented by the enetLTS R package \cite{enetLTS}.
Two different versions of enetLTS are used. 
One is nearly $\ell_1$ penalty (denoted by LTS in the tables) and the other is an optimal chosen elastic net penalty (denoted by LTS.full). 
The tuning parameter $\lambda$ is selected via 5-fold cross-validation from a grid of 60 values in $[10^{-4}, 0.1]$. 

We still use three measures to compare the methods. 
MSE and FSL are defined in the same way as in the Gaussian linear regression case. 
The measure ME($\%$) is the percentage of misclassification error on the test data set, i.e., $\text{ME}=\frac{1}{100}\sum_{i=1}^{100}I(y_{i,\text{test}}\neq \hat{y}_{i,\text{test}})$. 
The results over 50 replicates are presented in Table~\ref{tab:logistic_identity} ($\bm \Sigma_X=\bm{I}_p$) and Table~\ref{tab:logistic_ar} ($\bm \Sigma_X=$AR(0.7)). 

In the absence of contamination ($\tau=0$), the standard Lasso generally achieves the lowest MSE, although $\hat{\bm{\theta}}$ and $\tilde{\bm{\theta}}$ remain competitive. However, as soon as contamination is introduced, the performance of Lasso degrades significantly, particularly under the $\mathcal{X}_2$, $\mathcal{XY}_1$ and $\mathcal{XY}_2$ settings where the active predictors are corrupted. 

The two enetLTS benchmarks exhibit unexpected instability in these high-dimensional settings; its MSE and PE often exceed those of even the non-robust Lasso. 
This suggests that existing robust classification methods may struggle when outliers are high-leverage points in a large-scale parameter space. 
In contrast, both MMD-based estimators ($\hat{\bm{\theta}}$ and $\tilde{\bm{\theta}}$) demonstrate remarkable resilience. 
Specifically, $\tilde{\bm{\theta}}$ consistently yields the lowest MSE and Prediction Error (PE) across nearly all contaminated scenarios.

With respect to feature selection, the penalized MMD framework ($\hat{\bm{\theta}}$) achieves the lowest FSL in the majority of settings, demonstrating its ability to distinguish true active predictors from noise even when those predictors are the source of contamination. 
As $p$ increases from 100 to 200, the performance gap between the proposed methods and the benchmarks further widens, particularly in the AR(0.7) covariance setting. 
Finally, we observe that the $O(n)$ reduced estimator $\tilde{\bm{\theta}}$ often provides superior predictive accuracy compared to the full $O(n^2)$ version in classification tasks, further supporting the practical utility of the local MMD approximation for high-dimensional classification.

\begin{table}[htbp]
\centering
\caption{Logistic Regression results for $\bm{\Sigma}_X = \bm{I}_p$ (Identity). The smallest mean in each row is in bold.}
\label{tab:logistic_identity}
\footnotesize
\setlength{\tabcolsep}{5pt}
\begin{tabular}{ccccccccc}
\toprule
$p$ & Metric & $\tau$ & $\mathcal{T}$ & Lasso & LTS & LTS.full & $\hat{\bm \theta}_n$ & $\tilde{\bm \theta}_n$ \\
\midrule
\multirow{27}{*}{100} & \multirow{9}{*}{MSE} & 0 & -- & \textbf{5.01} (0.97) & 11.2 (6.0) & 8.03 (2.5) & 5.45 (1.8) & 5.29 (1.3) \\
\cmidrule(lr){3-9}
& & \multirow{4}{*}{0.05} & $\mathcal{X}_1$ & \textbf{7.61} (1.5) & 11.5 (5.2) & 9.91 (3.7) & 7.83 (1.2) & 7.69 (1.4) \\
& & & $\mathcal{X}_2$ & 9.46 (0.31) & 14.9 (3.6) & 14.3 (6.0) & 9.48 (0.32) & \textbf{7.34} (1.5) \\
& & & $\mathcal{XY}_1$ & 9.36 (0.28) & 15.5 (6.1) & 13.4 (4.6) & 9.37 (0.29) & \textbf{6.78} (1.5) \\
& & & $\mathcal{XY}_2$ & 9.49 (0.30) & 14.9 (4.3) & 13.3 (2.7) & \textbf{9.47} (0.29) & 8.17 (1.9) \\
\cmidrule(lr){3-9}
& & \multirow{4}{*}{0.1} & $\mathcal{X}_1$ & \textbf{7.02} (1.1) & 12.4 (4.3) & 10.6 (5.4) & 7.27 (1.3) & 7.60 (1.4) \\
& & & $\mathcal{X}_2$ & 9.46 (0.25) & 15.0 (3.9) & 13.3 (3.6) & 9.52 (0.40) & \textbf{7.27} (1.3) \\
& & & $\mathcal{XY}_1$ & 9.48 (0.31) & 15.9 (6.0) & 13.5 (3.2) & 9.49 (0.31) & \textbf{7.46} (1.7) \\
& & & $\mathcal{XY}_2$ & 9.48 (0.37) & 16.8 (6.1) & 15.2 (5.1) & \textbf{9.46} (0.30) & 8.32 (1.1) \\
\cmidrule(lr){2-9}
& \multirow{9}{*}{FSL} & 0 & -- & 17.2 (6.2) & 18.1 (5.5) & 36.7 (16.8) & \textbf{9.53} (3.6) & 17.3 (13.5) \\
\cmidrule(lr){3-9}
& & \multirow{4}{*}{0.05} & $\mathcal{X}_1$ & 13.6 (6.5) & 17.5 (5.5) & 37.1 (13.9) & \textbf{9.30} (3.6) & 20.7 (18.9) \\
& & & $\mathcal{X}_2$ & 10.5 (2.7) & 21.8 (7.6) & 45.5 (17.7) & \textbf{9.90} (1.6) & 21.9 (19.2) \\
& & & $\mathcal{XY}_1$ & 10.1 (2.6) & 21.9 (8.1) & 44.3 (19.9) & \textbf{9.50} (1.3) & 19.2 (16.5) \\
& & & $\mathcal{XY}_2$ & 11.0 (4.2) & 23.8 (7.0) & 41.9 (15.8) & \textbf{9.67} (1.1) & 26.1 (22.9) \\
\cmidrule(lr){3-9}
& & \multirow{4}{*}{0.1} & $\mathcal{X}_1$ & 15.2 (7.3) & 18.7 (6.0) & 30.5 (15.8) & \textbf{8.63} (2.9) & 20.2 (19.3) \\
& & & $\mathcal{X}_2$ & 10.5 (4.0) & 21.6 (7.5) & 44.9 (15.3) & \textbf{10.1} (3.7) & 16.3 (12.0) \\
& & & $\mathcal{XY}_1$ & 10.6 (3.5) & 24.3 (7.6) & 44.1 (15.3) & \textbf{9.93} (1.8) & 18.2 (16.4) \\
& & & $\mathcal{XY}_2$ & 12.6 (5.4) & 25.0 (7.2) & 42.0 (16.0) & \textbf{10.1} (1.9) & 23.9 (17.0) \\
\cmidrule(lr){2-9}
& \multirow{9}{*}{ME (\%)} & 0 & -- & 26.9 (5.4) & 37.6 (6.7) & 34.0 (6.5) & 28.1 (6.7) & \textbf{26.7} (5.1) \\
\cmidrule(lr){3-9}
& & \multirow{4}{*}{0.05} & $\mathcal{X}_1$ & \textbf{32.8} (5.4) & 40.8 (6.0) & 37.4 (6.0) & 34.3 (4.2) & 33.4 (5.0) \\
& & & $\mathcal{X}_2$ & 41.4 (6.5) & 50.6 (4.4) & 51.6 (4.7) & 42.2 (7.5) & \textbf{33.0} (6.5) \\
& & & $\mathcal{XY}_1$ & 40.7 (7.0) & 51.4 (5.2) & 51.4 (4.7) & 41.1 (8.0) & \textbf{32.5} (8.3) \\
& & & $\mathcal{XY}_2$ & 40.3 (6.1) & 48.7 (4.5) & 50.1 (6.6) & 41.9 (8.2) & \textbf{33.3} (6.9) \\
\cmidrule(lr){3-9}
& & \multirow{4}{*}{0.1} & $\mathcal{X}_1$ & \textbf{31.0} (4.9) & 40.5 (5.4) & 38.4 (5.6) & 32.1 (6.5) & 31.1 (5.7) \\
& & & $\mathcal{X}_2$ & 38.7 (5.8) & 49.6 (5.5) & 50.6 (7.3) & 40.3 (8.4) & \textbf{32.1} (6.1) \\
& & & $\mathcal{XY}_1$ & 40.1 (5.7) & 50.2 (6.9) & 49.6 (4.7) & 40.8 (7.1) & \textbf{32.7} (8.6) \\
& & & $\mathcal{XY}_2$ & 41.1 (6.3) & 49.2 (4.9) & 50.6 (5.4) & 43.0 (9.2) & \textbf{36.0} (8.6) \\
\midrule
\multirow{27}{*}{200} & \multirow{9}{*}{MSE} & 0 & -- & \textbf{6.77} (1.4) & 10.9 (3.0) & 8.87 (2.3) & 7.17 (1.4) & 7.28 (1.5) \\
\cmidrule(lr){3-9}
& & \multirow{4}{*}{0.05} & $\mathcal{X}_1$ & \textbf{7.71} (1.0) & 11.0 (2.6) & 10.0 (2.6) & 8.20 (1.2) & 8.00 (1.1) \\
& & & $\mathcal{X}_2$ & 9.45 (0.31) & 14.5 (4.1) & 12.6 (3.4) & 9.48 (0.35) & \textbf{7.77} (1.5) \\
& & & $\mathcal{XY}_1$ & 9.44 (0.22) & 14.9 (4.4) & 12.8 (3.1) & 9.44 (0.23) & \textbf{7.65} (1.2) \\
& & & $\mathcal{XY}_2$ & 9.34 (0.26) & 14.1 (3.7) & \textbf{11.3} (1.3) & 9.34 (0.28) & 8.39 (1.3) \\
\cmidrule(lr){3-9}
& & \multirow{4}{*}{0.1} & $\mathcal{X}_1$ & \textbf{8.15} (1.1) & 12.5 (4.4) & 9.66 (3.5) & 8.31 (1.1) & 8.12 (0.89) \\
& & & $\mathcal{X}_2$ & 9.39 (0.21) & 13.7 (3.1) & 11.6 (1.5) & 9.39 (0.22) & \textbf{7.63} (0.97) \\
& & & $\mathcal{XY}_1$ & 9.59 (0.40) & 16.5 (6.7) & 13.3 (4.2) & 9.57 (0.31) & \textbf{8.31} (1.6) \\
& & & $\mathcal{XY}_2$ & \textbf{9.52} (0.29) & 15.3 (3.9) & 12.7 (2.6) & 9.55 (0.34) & 9.44 (1.5) \\
\cmidrule(lr){2-9}
& \multirow{9}{*}{FSL} & 0 & -- & 15.2 (6.6) & 19.9 (7.0) & 38.7 (28.0) & \textbf{8.53} (3.6) & 12.5 (10.1) \\
\cmidrule(lr){3-9}
& & \multirow{4}{*}{0.05} & $\mathcal{X}_1$ & 15.7 (8.3) & 19.4 (8.0) & 40.1 (24.4) & \textbf{10.2} (3.8) & 21.0 (20.6) \\
& & & $\mathcal{X}_2$ & 11.5 (4.0) & 26.8 (8.4) & 47.8 (27.3) & \textbf{10.4} (2.3) & 25.6 (22.4) \\
& & & $\mathcal{XY}_1$ & 10.6 (4.2) & 26.3 (9.2) & 50.6 (23.7) & \textbf{9.60} (0.93) & 18.7 (12.7) \\
& & & $\mathcal{XY}_2$ & 10.4 (3.3) & 24.7 (8.6) & 59.6 (24.5) & \textbf{9.57} (0.94) & 27.5 (24.3) \\
\cmidrule(lr){3-9}
& & \multirow{4}{*}{0.1} & $\mathcal{X}_1$ & 13.9 (7.9) & 24.6 (6.8) & 45.5 (31.1) & \textbf{9.17} (2.5) & 26.3 (22.1) \\
& & & $\mathcal{X}_2$ & 11.4 (5.1) & 21.7 (7.5) & 49.2 (25.8) & \textbf{10.1} (2.6) & 23.3 (19.4) \\
& & & $\mathcal{XY}_1$ & 10.6 (4.2) & 27.2 (11.3) & 59.9 (36.0) & \textbf{9.67} (1.3) & 21.5 (22.3) \\
& & & $\mathcal{XY}_2$ & 11.8 (5.4) & 28.3 (8.1) & 57.7 (28.7) & \textbf{10.7} (3.1) & 31.2 (31.5) \\
\cmidrule(lr){2-9}
& \multirow{9}{*}{ME (\%)} & 0 & -- & \textbf{30.9} (6.5) & 38.8 (5.9) & 36.3 (5.4) & 32.6 (6.1) & 31.4 (6.3) \\
\cmidrule(lr){3-9}
& & \multirow{4}{*}{0.05} & $\mathcal{X}_1$ & \textbf{34.2} (6.4) & 39.9 (7.7) & 40.9 (7.6) & 37.0 (7.6) & 36.2 (5.9) \\
& & & $\mathcal{X}_2$ & 40.7 (5.2) & 49.9 (5.7) & 51.2 (7.4) & 41.7 (6.6) & \textbf{34.3} (6.8) \\
& & & $\mathcal{XY}_1$ & 39.4 (5.2) & 51.9 (5.1) & 49.5 (6.1) & 40.7 (7.4) & \textbf{32.7} (5.9) \\
& & & $\mathcal{XY}_2$ & 40.5 (6.2) & 49.4 (5.1) & 51.9 (5.2) & 39.8 (5.0) & \textbf{35.5} (6.6) \\
\cmidrule(lr){3-9}
& & \multirow{4}{*}{0.1} & $\mathcal{X}_1$ & \textbf{36.6} (9.4) & 42.7 (8.0) & 42.2 (7.0) & 36.7 (10.0) & 38.9 (8.1) \\
& & & $\mathcal{X}_2$ & 40.2 (7.5) & 51.0 (4.8) & 51.8 (6.5) & 40.4 (7.7) & \textbf{34.7} (8.3) \\
& & & $\mathcal{XY}_1$ & 41.6 (6.9) & 51.8 (4.9) & 51.1 (4.5) & 43.4 (8.7) & \textbf{36.9} (8.8) \\
& & & $\mathcal{XY}_2$ & 45.1 (6.8) & 48.9 (5.8) & 47.5 (4.4) & 47.7 (7.7) & \textbf{20.7} (6.8) \\
\bottomrule
\end{tabular}
\end{table}

\begin{table}[htbp]
\centering
\caption{Logistic Regression results for $\bm{\Sigma}_X = \text{AR}(0.7)$. The smallest mean in each row is in bold.}
\label{tab:logistic_ar}
\footnotesize
\setlength{\tabcolsep}{5pt}
\begin{tabular}{ccccccccc}
\toprule
$p$ & Metric & $\tau$ & $\mathcal{T}$ & Lasso & LTS & LTS.full & $\hat{\bm \theta}_n$ & $\tilde{\bm \theta}_n$ \\
\midrule
\multirow{27}{*}{100} & \multirow{9}{*}{MSE} & 0 & -- & \textbf{4.87} (1.2) & 10.2 (7.6) & 7.09 (2.6) & 5.20 (1.4) & 5.74 (1.7) \\
\cmidrule(lr){3-9}
& & \multirow{4}{*}{0.05} & $\mathcal{X}_1$ & \textbf{5.37} (1.0) & 9.95 (5.1) & 8.48 (3.9) & 5.38 (0.58) & 5.94 (0.99) \\
& & & $\mathcal{X}_2$ & 9.74 (0.45) & 20.2 (9.3) & 14.6 (4.7) & 9.69 (0.35) & \textbf{6.11} (1.1) \\
& & & $\mathcal{XY}_1$ & 9.70 (0.42) & 20.4 (9.3) & 14.4 (3.7) & 9.72 (0.30) & \textbf{6.09} (1.5) \\
& & & $\mathcal{XY}_2$ & 9.66 (0.29) & 15.3 (3.6) & 14.8 (6.7) & 9.68 (0.32) & \textbf{7.18} (0.90) \\
\cmidrule(lr){3-9}
& & \multirow{4}{*}{0.1} & $\mathcal{X}_1$ & \textbf{5.57} (0.85) & 12.6 (8.1) & 8.52 (4.0) & 5.55 (0.91) & 6.53 (1.3) \\
& & & $\mathcal{X}_2$ & 9.79 (0.53) & 18.2 (7.3) & 16.6 (7.3) & 9.80 (0.38) & \textbf{6.11} (1.4) \\
& & & $\mathcal{XY}_1$ & 9.77 (0.39) & 17.4 (4.7) & 14.0 (3.2) & 9.73 (0.33) & \textbf{7.13} (1.5) \\
& & & $\mathcal{XY}_2$ & 9.79 (0.38) & 18.5 (8.8) & 15.1 (4.2) & 9.80 (0.36) & \textbf{8.39} (1.4) \\
\cmidrule(lr){2-9}
& \multirow{9}{*}{FSL} & 0 & -- & 12.3 (5.2) & 12.5 (4.3) & 18.5 (11.0) & \textbf{6.37} (2.7) & 10.6 (7.9) \\
\cmidrule(lr){3-9}
& & \multirow{4}{*}{0.05} & $\mathcal{X}_1$ & 12.1 (5.3) & 13.5 (4.8) & 18.8 (7.7) & \textbf{7.53} (3.1) & 10.3 (6.9) \\
& & & $\mathcal{X}_2$ & 13.4 (5.8) & 24.5 (7.1) & 43.8 (16.0) & \textbf{10.5} (1.9) & 7.03 (7.9) \\
& & & $\mathcal{XY}_1$ & 11.7 (4.6) & 25.9 (6.1) & 44.3 (14.0) & \textbf{10.5} (1.6) & 10.4 (11.4) \\
& & & $\mathcal{XY}_2$ & 12.2 (3.9) & 22.1 (5.2) & 41.6 (15.4) & \textbf{10.8} (2.7) & 8.33 (6.6) \\
\cmidrule(lr){3-9}
& & \multirow{4}{*}{0.1} & $\mathcal{X}_1$ & 13.4 (5.4) & 14.8 (6.4) & 18.1 (6.5) & \textbf{7.40} (2.8) & 10.7 (7.7) \\
& & & $\mathcal{X}_2$ & 12.3 (4.1) & 23.3 (7.2) & 44.0 (16.7) & \textbf{11.2} (3.2) & 6.57 (4.3) \\
& & & $\mathcal{XY}_1$ & 11.7 (3.1) & 24.9 (6.8) & 41.0 (15.3) & \textbf{10.5} (1.7) & 6.03 (3.7) \\
& & & $\mathcal{XY}_2$ & 12.1 (3.9) & 23.2 (6.7) & 43.9 (15.3) & \textbf{10.8} (2.1) & 14.0 (14.2) \\
\cmidrule(lr){2-9}
& \multirow{9}{*}{ME (\%)} & 0 & -- & \textbf{13.3} (3.4) & 16.2 (4.7) & 17.1 (5.2) & 13.8 (3.4) & 14.5 (3.2) \\
\cmidrule(lr){3-9}
& & \multirow{4}{*}{0.05} & $\mathcal{X}_1$ & 14.3 (4.5) & 17.3 (5.5) & 16.5 (4.6) & 14.2 (4.5) & \textbf{14.1} (3.9) \\
& & & $\mathcal{X}_2$ & 43.1 (8.9) & 43.3 (9.3) & 42.1 (9.2) & 43.7 (8.6) & \textbf{14.0} (4.8) \\
& & & $\mathcal{XY}_1$ & 42.9 (7.0) & 46.0 (7.6) & 42.8 (7.7) & 43.8 (6.5) & \textbf{13.3} (4.2) \\
& & & $\mathcal{XY}_2$ & 41.2 (5.6) & 44.5 (6.1) & 45.0 (6.8) & 42.7 (6.7) & \textbf{16.1} (4.4) \\
\cmidrule(lr){3-9}
& & \multirow{4}{*}{0.1} & $\mathcal{X}_1$ & \textbf{12.3} (3.8) & 16.3 (6.5) & 17.0 (5.3) & 12.9 (3.9) & 12.9 (3.8) \\
& & & $\mathcal{X}_2$ & 43.1 (7.0) & 45.9 (6.8) & 47.5 (6.0) & 43.5 (7.8) & \textbf{12.7} (3.1) \\
& & & $\mathcal{XY}_1$ & 44.3 (7.0) & 46.9 (6.7) & 45.5 (8.2) & 44.9 (7.2) & \textbf{17.3} (12.8) \\
& & & $\mathcal{XY}_2$ & 45.9 (7.4) & 47.0 (7.9) & 46.3 (4.2) & 47.5 (9.2) & \textbf{20.6} (6.4) \\
\midrule
\multirow{27}{*}{200} & \multirow{9}{*}{MSE} & 0 & -- & \textbf{5.31} (1.3) & 8.64 (4.1) & 7.61 (4.1) & 5.38 (1.2) & 5.71 (1.4) \\
\cmidrule(lr){3-9}
& & \multirow{4}{*}{0.05} & $\mathcal{X}_1$ & \textbf{5.59} (0.67) & 8.21 (4.2) & 7.34 (2.3) & 5.81 (1.1) & 6.06 (1.1) \\
& & & $\mathcal{X}_2$ & 9.68 (0.40) & 16.3 (5.9) & 13.1 (3.9) & 9.65 (0.26) & \textbf{6.84} (1.5) \\
& & & $\mathcal{XY}_1$ & 9.71 (0.36) & 18.5 (11.7) & 13.0 (3.6) & 9.72 (0.36) & \textbf{6.48} (1.1) \\
& & & $\mathcal{XY}_2$ & 9.69 (0.37) & 15.3 (3.5) & 13.3 (4.6) & 9.70 (0.39) & \textbf{7.76} (1.1) \\
\cmidrule(lr){3-9}
& & \multirow{4}{*}{0.1} & $\mathcal{X}_1$ & 6.23 (2.7) & 9.96 (5.2) & 7.60 (2.0) & 6.27 (1.3) & 6.81 (1.2) \\
& & & $\mathcal{X}_2$ & 9.70 (0.33) & 16.5 (6.2) & 13.9 (4.0) & 9.75 (0.37) & \textbf{6.59} (0.93) \\
& & & $\mathcal{XY}_1$ & 9.68 (0.32) & 19.3 (10.5) & 13.9 (3.8) & 9.71 (0.32) & \textbf{6.67} (1.2) \\
& & & $\mathcal{XY}_2$ & 9.78 (0.43) & 16.6 (5.9) & 13.9 (3.8) & 9.77 (0.32) & \textbf{8.19} (0.78) \\
\cmidrule(lr){2-9}
& \multirow{9}{*}{FSL} & 0 & -- & 14.8 (5.8) & 15.7 (5.4) & 20.4 (8.6) & \textbf{7.93} (3.8) & 10.6 (7.4) \\
\cmidrule(lr){3-9}
& & \multirow{4}{*}{0.05} & $\mathcal{X}_1$ & 15.1 (7.7) & 15.5 (5.6) & 20.6 (10.2) & \textbf{8.47} (4.0) & 12.3 (8.3) \\
& & & $\mathcal{X}_2$ & 12.0 (4.0) & 24.8 (7.8) & 61.5 (28.8) & \textbf{10.1} (0.93) & 7.07 (3.4) \\
& & & $\mathcal{XY}_1$ & 11.9 (4.0) & 26.2 (8.7) & 53.6 (21.6) & \textbf{10.4} (2.1) & 7.60 (7.0) \\
& & & $\mathcal{XY}_2$ & 12.0 (3.8) & 25.9 (8.0) & 73.6 (30.5) & \textbf{10.3} (1.8) & 7.23 (8.1) \\
\cmidrule(lr){3-9}
& & \multirow{4}{*}{0.1} & $\mathcal{X}_1$ & 15.7 (6.3) & 17.3 (6.2) & 25.7 (18.7) & \textbf{8.30} (3.5) & 9.20 (4.6) \\
& & & $\mathcal{X}_2$ & 11.9 (3.0) & 26.4 (9.1) & 56.0 (27.5) & \textbf{10.4} (1.4) & 6.37 (3.6) \\
& & & $\mathcal{XY}_1$ & 11.3 (2.6) & 28.9 (8.9) & 51.6 (23.5) & \textbf{10.4} (2.1) & 7.17 (4.8) \\
& & & $\mathcal{XY}_2$ & 12.4 (6.9) & 29.3 (7.6) & 55.6 (28.3) & \textbf{10.3} (2.3) & 10.7 (7.9) \\
\cmidrule(lr){2-9}
& \multirow{9}{*}{ME (\%)} & 0 & -- & \textbf{14.1} (4.3) & 18.6 (4.9) & 17.0 (4.1) & 14.2 (3.3) & 14.8 (3.4) \\
\cmidrule(lr){3-9}
& & \multirow{4}{*}{0.05} & $\mathcal{X}_1$ & \textbf{14.7} (3.4) & 17.2 (4.8) & 18.5 (5.1) & 14.9 (4.8) & 15.4 (4.6) \\
& & & $\mathcal{X}_2$ & 44.5 (6.5) & 44.8 (9.2) & 44.4 (7.7) & 46.0 (6.6) & \textbf{16.9} (10.1) \\
& & & $\mathcal{XY}_1$ & 44.5 (6.6) & 45.5 (9.4) & 42.4 (10.4) & 45.2 (7.5) & \textbf{15.0} (9.0) \\
& & & $\mathcal{XY}_2$ & 43.5 (6.5) & 44.5 (8.1) & 45.4 (7.9) & 45.5 (7.0) & \textbf{19.0} (10.9) \\
\cmidrule(lr){3-9}
& & \multirow{4}{*}{0.1} & $\mathcal{X}_1$ & 14.4 (3.9) & 17.1 (4.2) & 18.5 (4.4) & 14.9 (4.2) & 15.9 (3.6) \\
& & & $\mathcal{X}_2$ & 40.7 (6.1) & 45.9 (5.4) & 47.4 (5.8) & 43.0 (7.3) & \textbf{15.3} (3.6) \\
& & & $\mathcal{XY}_1$ & 44.4 (7.5) & 47.3 (5.0) & 47.8 (6.2) & 46.2 (7.4) & \textbf{15.9} (7.3) \\
& & & $\mathcal{XY}_2$ & 45.1 (6.8) & 48.9 (5.8) & 47.5 (4.4) & 47.7 (7.7) & \textbf{20.7} (6.8) \\
\bottomrule
\end{tabular}
\end{table}

\section{Real Data Applications}

%In this section, we evaluate the performance of the proposed estimators on two real-world datasets to demonstrate their practical utility in high-dimensional settings.

\begin{figure}[htbp]
\centering
\begin{subfigure}[b]{0.4\textwidth}
\includegraphics[width=\textwidth]{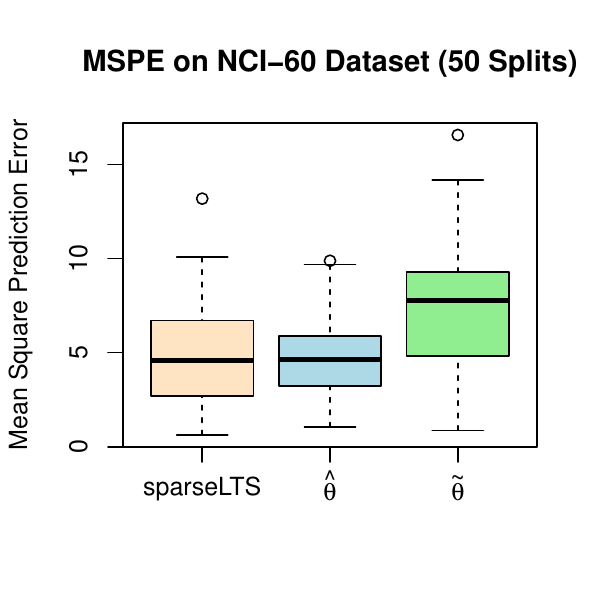}
\caption{MESPEs for NCI-60 Example.\label{fig:boxplot_nci60}}
\end{subfigure}
\begin{subfigure}[b]{0.4\textwidth}
\includegraphics[width=\textwidth]{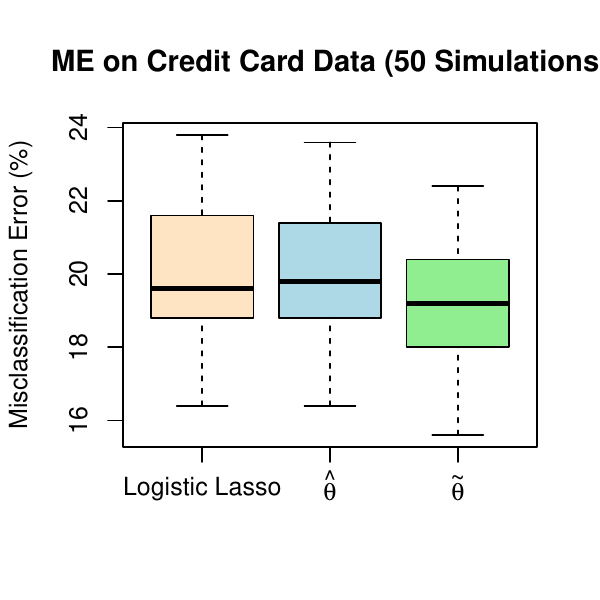}
\caption{MEs (\%) for Credit Example.\label{fig:boxplot_credit}}
\end{subfigure}
\caption{Results of the two real-data applications.}
\end{figure}

We first consider the NCI-60 cancer cell line panel \citep{reinhold2012cellminer}, which is a widely used dataset in bioinformatics for studying the relationship between gene expressions and protein levels. Following the example in the \texttt{robustHD} package \citep{alfons2021robusthd, alfons2013sparse}, we select the protein expression of the 92nd protein as the response variable. 
To create a high-dimensional regression scenario, we screen the gene expressions and select the 100 genes with the highest robustly estimated correlations (using Huber correlation) with the response. 
The resulting dataset contains $n=59$ observations and $p=100$ predictors.
To compare the predictive performance of the proposed estimators ($\hat{\bm{\theta}}_n$ and $\tilde{\bm{\theta}}_n$) against the sparse least trimmed squares (\texttt{sparseLTS}) estimator, we perform a random splitting experiment. 
The data is randomly partitioned into a training set ($n_1=45$) and a testing set ($n_2=14$). 
This process is repeated 50 times. For each split, we select the tuning parameters via 5-fold cross-validation on the training data and calculate the Mean Square Prediction Error (MSPE) on the independent test set. 
The distribution of the MSPE across the 50 replicates is shown in Figure~\ref{fig:boxplot_nci60}. 
The proposed MMD-based estimator $\hat{\bm \theta}_n$ exhibits lower median MSPE and higher stability (narrower interquartile range) compared to \texttt{sparseLTS}, but $\tilde{\bm \theta}_n$ performs worse than the two. 

Next, we apply our methodology to the ``Default of Credit Card Clients'' dataset \citep{yeh2009comparisons}, containing 30,000 observations and 23 predictors. 
While this dataset has a larger sample size ($n > p$), it allows us to evaluate the computational efficiency of our $O(n)$ approximate estimator $\tilde{\bm \theta}$ and its robustness to class imbalance and noise in a ``large-scale'' setting. Even when $p$ is not larger than $n$, the sparsity induced by the $\ell_1$ penalty remains crucial for identifying the most influential predictors in the presence of contamination.
We performed 50 random splits where a training set of $n=2,000$ was sampled using stratified sampling to preserve the original class ratio, and a test set of $m=500$ was sampled randomly from the remaining observations. 
Variable selection and estimation were performed using 23 predictors, with selective standardization applied to the continuous variables (limit amount, age, and payment/bill records). 
We compare the classification performance of our proposed $\hat{\bm{\theta}}_n$ and $\tilde{\bm{\theta}}_n$ estimators with the standard logistic Lasso. 
The misclassification errors, summarized in Figure~\ref{fig:boxplot_credit}, indicate that the MMD-based estimator $\tilde{\bm \theta}$ provide a more robust classification boundary than the standard Lasso, yielding lower average prediction error and significantly reduced variability across different data splits. 
This suggests that MMD’s joint-distribution perspective is particularly advantageous for financial datasets characterized by noise and class imbalance.

\section{Conclusion}\label{sec:end}

In this work, we have extended the robust MMD-based estimation framework to high-dimensional Generalized Linear Models by incorporating sparsity-inducing penalties. By integrating an $\ell_1$ penalty with the MMD distance, we provide a unified approach that achieves simultaneous robustness to outliers and effective variable selection. Our use of the ADMM framework and the $O(n)$ local approximation makes the method computationally viable for high-dimensional settings where $p > n$.

The simulation results confirm that our proposed estimators, $\hat{\theta}_n$ and $\tilde{\theta}_n$, offer a unique universal' robustness profile. While specialized estimators like the Huber-Lasso excel in response-only outlier scenarios, the MMD-based framework proves significantly more resilient when facing high-leverage points and heavy-tailed error distributions ($t_5$ and Laplace) simultaneously. Crucially, our methods consistently achieved the lowest False Selection Loss (FSL) across most contaminated scenarios, demonstrating that the MMD loss function is less likely to be 'distracted' by outliers during the feature selection process—a vital property for high-dimensional discovery tasks.

Our real-data applications further highlight the practical utility of this framework. In the NCI-60 cancer cell line study, the MMD estimator $\hat{\theta}_n$ provided higher predictive stability and lower median MSPE than the sparseLTS benchmark. In the large-scale credit card default analysis, the $O(n)$ approximate estimator $\tilde{\theta}_n$ proved to be highly efficient, delivering a more robust classification boundary than the standard logistic Lasso in the presence of noisy financial data and class imbalance.

Future research could investigate the theoretical oracle properties of these penalized MMD estimators and explore automated, data-driven methods for bandwidth selection to further enhance the model's 'universal' applicability across different GLM kernels. Additionally, extending this framework to group-sparse penalties for multi-modal data integration remains a promising direction.

\bibliography{ref.bib}% common bib file

\newpage 
\phantomsection\label{supplementary-material}
\bigskip

\setcounter{page}{1}
\setcounter{figure}{0}
\setcounter{table}{0}
\setcounter{theorem}{0}

\makeatletter 
\renewcommand{\thefigure}{S\@arabic\c@figure}
\renewcommand{\thetable}{S\@arabic\c@table}
\renewcommand{\theproposition}{S\@arabic\c@proposition}
\renewcommand{\thetheorem}{S\@arabic\c@theorem}
\makeatother

\begin{center}

{\large\bf SUPPLEMENTARY MATERIAL}

\end{center}

The supplementary material contains derivations of the MMD loss function and their special cases for Gaussian linear and logistic regression models. 

\section*{Derivations of $\textcircled{1}$ and $\textcircled{2}$}

\begin{align*}
\textcircled{1}&=\int K(\bm z, \bm z')P_{\bm \theta}^n(\dd \bm z)P_{\bm \theta}^n(\dd \bm z') \\
&=\int_x K_x(\bm x,\bm x')\int_y K_y(y, y')P_{g(\bm \theta, \bm x)}(\dd y)P_{g(\bm \theta, \bm x')}(\dd y') \hat{P}^n_x(\dd \bm x)\hat{P}^n_x(\dd \bm x')\\
&=\frac{1}{n^2}\sum_{i,j=1}^n K_x(\bm x_i, \bm x_j)\int_y K_y(y,y')p_{g(\bm \theta,\bm x_i)}(y)p_{g(\bm \theta,\bm x_j)}(y')\dd y \dd y'\\
&=\frac{1}{n^2}\sum_{i,j=1}^n K_x(\bm x_i, \bm x_j)\mathbb{E}_{Y\sim P_{g(\bm \theta,\bm x_i)}, Y'\sim P_{g(\bm \theta, \bm x_j)}}[K_y(Y,Y')].\\
\textcircled{2}&=-\frac{2}{n}\sum_{i=1}^n\int K_x(\bm x, \bm x_i)K_y(y, y_i) p_{g(\bm \theta, \bm x)}(y)\dd y \hat{P}^n_x(\dd \bm x)\\
&=-\frac{2}{n^2}\sum_{i,j=1}^n K_x(\bm x_i,\bm x_j) \int K_y(y, y_i)p_{g(\bm \theta, \bm x_j)}(y)\dd y.
\end{align*}

\section*{Derivations for Gaussian Linear Model}

Now we derive $l(\bm \theta, \bm x_i, y_i)$ and $l(\bm \theta, \bm x_i, \bm x_j, y_i)$.
These following terms are needed.
\begin{align*}
& \mathbb{E}_{Y\sim N(\mu,\sigma^2)}[K_y(Y, y_0)]=\frac{1}{\sqrt{2\pi}\sigma}\int \exp\left(-\frac{(y-y_0)^2}{2h_y^2}\right)\exp\left(-\frac{(y-\mu)^2}{2\sigma^2}\right)\dd y\\
=&\frac{1}{\sigma}\exp\left(-\frac{1}{2}\left[\frac{\mu^2}{\sigma^2}+\frac{y_0^2}{h_y^2}\right]\right)\left(\frac{1}{\sigma^2}+\frac{1}{h_y^2}\right)^{-1/2}\times \\
& \int \frac{1}{\sqrt{2\pi}}\frac{1}{\left(\frac{1}{\sigma^2}+\frac{1}{h_y^2}\right)^{-1/2}} \exp\left(
-\frac{1}{2\left(\frac{1}{\sigma^2}+\frac{1}{h_y^2}\right)^{-1}}\left[y-\left(\frac{\mu}{\sigma^2}+\frac{y_0}{h_y^2}\right)\left(\frac{1}{\sigma^2}+\frac{1}{h_y^2}\right)^{-1}\right]^2\right)\dd y \\
& \times \exp\left(\frac{1}{2}\left(\frac{\mu}{\sigma^2}+\frac{y_0}{h_y^2}\right)^2\left(\frac{1}{\sigma^2}+\frac{1}{h_y^2}\right)^{-1}\right)\\
=&\frac{h_y}{\sqrt{\sigma^2+h_y^2}}\exp\left(-\frac{1}{2}\frac{(y_0-\mu)^2}{\sigma^2+h_y^2}\right)
\end{align*}
Applying it to $\mathbb{E}_{Y\sim P_{g(\bm \theta, \bm x_j)}}[K_y(Y, y_i)]$, $\mathbb{E}_{Y,Y'\sim P_{g(\bm \theta,\bm x_i)}}[K_y(Y,Y')]$, and $\mathbb{E}_{Y\sim P_{g(\bm \theta,\bm x_i)},Y'\sim P_{g(\bm \theta,\bm x_j)}}[K_y(Y,Y')]$, we can obtain the following results. 
\begin{align*}
&\mathbb{E}_{Y\sim P_{g(\bm \theta, \bm x_j)}}[K_y(Y, y_i)]=\mathbb{E}_{Y\sim N(\bm x_j^\top \bm \theta, \sigma^2)}[K_y(Y, y_i)]=\frac{h_y}{\sqrt{\sigma^2+h_y^2}}\exp\left(-\frac{1}{2}\frac{(y_i-\bm x_j^\top \bm \theta)^2}{\sigma^2+h_y^2}\right),\\
& \mathbb{E}_{Y,Y'\sim P_{g(\bm \theta,\bm x_i)}}[K_y(Y,Y')]=\mathbb{E}_{Y, Y'\sim N(\bm x_j^\top \bm \theta, \sigma^2)}[K_y(Y, Y')]=\left(2\frac{\sigma^2}{h_y^2}+1\right)^{-1/2}=\frac{h_y}{\sqrt{2\sigma^2+h_y^2}},\\
& \mathbb{E}_{Y\sim P_{g(\bm \theta,\bm x_i)},Y'\sim P_{g(\bm \theta,\bm x_j)}}[K_y(Y,Y')]=\mathbb{E}_{Y\sim N(\bm x_i^\top \bm \theta,\sigma^2),Y'\sim N(\bm x_j^\top \bm \theta, \sigma^2)}[K_y(Y,Y')]\\
=&\frac{h_y}{\sqrt{2\sigma^2+h_y^2}}\exp\left(-\frac{1}{2}\frac{\bm \theta^\top (\bm x_j-\bm x_i)(\bm x_j-\bm x_i)^\top \bm \theta}{2\sigma^2+h_y^2}\right).
\end{align*}
Replacing these terms in $l(\bm \theta, \bm x_i, \bm x_j, y_i)$ and $\tilde{l}(\bm \theta, \bm x_i, y_i)$ leads to \eqref{eq:linearv1} and \eqref{eq:linearv2}.

\section*{Derivations for Binary Logistic Model}

\begin{align*}
l(\bm \theta, \bm x_i,\bm x_j,y_i)&=\sum_{y,y'=0}^1 0.5 h_y(1-h_y)^{|y-y'|}p_{g(\bm \theta, \bm x_i)}(y)p_{g(\bm \theta, \bm x_j)}(y')\\
&-2\sum_{y=0}^1 0.5 h_y(1-h_y)^{|y-y_i|}p_{g(\bm \theta, \bm x_j)}(y)\\
&=0.5h_y\left[\pi_i\pi_j+(1-\pi_i)(1-\pi_j)+(1-h_y)\pi_i(1-\pi_j)+(1-h_y)(1-\pi_i)\pi_j\right]\\
&-h_y\left[(1-h_y)^{1-y_i}\pi_j+(1-h_y)^{y_i}(1-\pi_j)\right]\\
&=0.5h_y\left[1-h_y(\pi_i+\pi_j)+2h_y\pi_i\pi_j\right]-h_y\left[(1-h_y)^{1-y_i}\pi_j+(1-h_y)^{y_i}(1-\pi_j)\right].
\end{align*}
\begin{align*}
l(\bm \theta, \bm x_i,y_i)&=\sum_{y,y'=0}^1 0.5 h_y(1-h_y)^{|y-y'|}p_{g(\bm \theta, \bm x_i)}(y)p_{g(\bm \theta, \bm x_i)}(y')\\
&-2\sum_{y=0}^1 0.5h_y(1-h_y)^{|y-y_i|}p_{g(\bm \theta,\bm x_i)}\\
&=0.5h_y\left[\pi_i^2+(1-\pi_i)^2+(1-h_y)\pi_i(1-\pi_i)+(1-h_y)(1-\pi_i)\pi_i\right]\\
&-h_y\left[(1-h_y)^{1-y_i}\pi_i+(1-h_y)^{y_i}(1-\pi_i)\right]\\
&=\left\{
\begin{array}{ll}
\text{if }y_i=0, & h_y^2\pi_i^2-0.5h_y\\
\text{if }y_i=1, & h_y^2(1-\pi_i)^2-0.5h_y
\end{array}\right.
=h_y^2\pi_i^{2(1-y_i)}(1-\pi_i)^{2y_i}-0.5h_y.
\end{align*}

\end{document}